\documentclass[a4paper,11pt]{article}
\pdfoutput=1
\usepackage{jheppub}  
\usepackage[english]{babel}
\usepackage[utf8x]{inputenc}
\usepackage{amsmath}
\usepackage{amssymb}
\usepackage[section]{placeins}
\usepackage{graphicx}
\usepackage{units}

\newcommand{\SecRef}[1]{section~\ref{#1}}
\newcommand{\FigRef}[1]{figure~\ref{#1}}
\newcommand{\pt}{p_\perp}
\newcommand{\kt}{k_\perp}
\newcommand{\qt}{q_\perp}
\newcommand{\alphas}{\alpha_\text{s}}
\newcommand{\ti}{T_\text{i}}
\newcommand{\taui}{\tau_\text{i}}
\newcommand{\tcut}{t_\text{c}}
\newcommand{\tc}{T_\text{c}}
\renewcommand{\d}{\text{d}}

\preprint{{\flushright CERN-PH-TH/2012-344\\ IPPP/12/92\\ DCPT/12/84 \\ MCnet-12-14\\}}

\title{A perturbative framework for jet quenching}

\author[a,b]{Korinna~C.~Zapp,}

\author[b]{Frank Krauss,}

\author[a]{Urs~A.~Wiedemann}

\affiliation[a]{Department of Physics, CERN, Theory Unit, CH-1211 Geneva 23}
\affiliation[b]{Institute for Particle Physics Phenomenology, Durham University, Durham\ DH1\,3LE, 
UK}

\emailAdd{korinna.zapp@cern.ch}

\abstract{
We present a conceptually new framework for describing jet evolution 
in the dense medium produced in ultra-relativistic nucleus-nucleus collisions 
using perturbative QCD and its implementation into the Monte Carlo event 
generator \textsc{Jewel}.  The rescattering of hard partons in the medium is 
modelled by infrared continued pQCD matrix elements supplemented with parton 
showers.  The latter approximate higher order real-emission matrix elements 
and thus generate medium-induced gluon emissions.  The interplay between 
different emissions is governed by their formation times.  The destructive 
interference between subsequent scattering processes, the non-Abelian version 
of the Landau--\-Pomeranchuk--\-Migdal effect, is also taken into account.  
In this way the complete radiation pattern is consistently treated in a 
uniform way.  Results obtained within this minimal and theoretically well 
constrained framework are compared with a variety of experimental data susceptible
to jet-quenching effects at both \textsc{RHIC} and the \textsc{LHC}.  Overall,
a good agreement between data and simulation is found.  This new
framework also allows to identify and quantify the dominant uncertainties in 
the simulation, and we show some relevant examples for this.
}

\begin{document}
\maketitle
\flushbottom

\section{Introduction}

High momentum transfer processes occur abundantly in hadronic interactions
at collider energies.  This is not only true for proton-proton collisions at 
the \textsc{LHC} and other similar collider experiments before it, but also 
for nucleus-nucleus collisions at \textsc{RHIC} and at the \textsc{LHC}.  
Such processes are well understood and calculable in perturbative QCD and 
they can be simulated with modern Monte Carlo event generators~\cite{
  Gleisberg:2008ta,Sjostrand:2007gs,Bahr:2008pv} using QCD matrix elements 
supplemented with parton showers~\cite{Buckley:2011ms}.  The latter resum the 
dominant collinear logarithms associated to the emission of additional partons 
to leading (and partly sub-leading) logarithmic accuracy, with many improvements
and reformulations in the past decades~\cite{
  Bengtsson:1986et,Marchesini:1987cf,Lonnblad:1992tz,Gieseke:2003rz,
  Nagy:2005aa,Giele:2007di,Schumann:2007mg,Dinsdale:2007mf,Winter:2007ye}.
In the present work, we discuss how such techniques can be used to formulate 
a general framework for the parton shower evolution in the presence of dense 
QCD matter produced in nucleus-nucleus collisions. 

\smallskip

Simple considerations, based on scale arguments and the uncertainty relation, 
indicate that final state parton showers evolve for up to several \unit{fm/c}, 
which is comparable to the transverse size of the dense QCD matter produced 
in the spatially extended overlap region of ultra-relativistic nucleus-nucleus 
collisions.  Therefore it can be expected that the QCD parton showers 
introduced and studied in the vacuum will be subjected to medium modifications 
in nucleus-nucleus collisions.  Data from \textsc{RHIC} and the \textsc{LHC} 
strongly support this view.  In particular, in comparison to standard 
perturbative benchmarks from $pp$--collisions, single inclusive 
hadron spectra at high transverse momentum are strongly suppressed, by up to a
factor of 5 at \textsc{RHIC}~\cite{Adare:2008qa,Abelev:2007ra}, and by up to 
a factor of 7 at the \textsc{LHC}~\cite{
  :2012eq,ATLAS-CONF-2012-120,CMS:2012aa}. 
Over the last two years, data on reconstructed jets in heavy-ion collisions 
have provided further, more differential information about such medium 
modifications of QCD radiation.  In particular, measurements of the di-jet 
asymmetry~\cite{Aad:2010bu,Chatrchyan:2011sx} and the inclusive jet 
suppression~\cite{:2012ch,:2012is} 
indicate that the interaction of parton showers with the surrounding QCD 
medium can dissipate a significant fraction of its total transverse energy 
to large angles, outside typical jet cones~\footnote{
  Reconstructed jets in heavy ion collisions are by now typically identified 
  through the anti-$\kt$ jet algorithm~\cite{Cacciari:2011ma}.
  The jet catchment areas that the anti-$\kt$ algorithm associates to jets 
  are approximately circular\cite{Cacciari:2008gn}, and we shall refer to them for simplicity 
  as 'cones' in the following.}. 
On the other hand, the fragmentation functions inside the jet cone appear to
experience only moderate medium modifications~\cite{
  Chatrchyan:2012gw,ATLAS-CONF-2012-115}. 

\smallskip

The modification of leading fragments has been anticipated more than two decades ago~\cite{Bjorken:1982tu,Braaten:1991jj,Gyulassy:1993hr}
and triggered the development of a large number of jet quenching models in
subsequent years.
For recent reviews on such models and their comparison to data from 
\textsc{RHIC} and the \textsc{LHC}, see refs.~\cite{
  d'Enterria:2009am,Wiedemann:2009sh,Renk:2012wi,CasalderreySolana:2012bp}.
In the following, we limit our discussion to highlighting the main similarities 
and the main differences between our proposal and some of these previous works. 
We start by recalling that there are essentially three qualitatively different 
analytically formulated approaches to medium-induced parton energy loss, all 
of which based on strong assumptions about kinematics and/or medium properties.  

Firstly, starting with refs.~\cite{Bjorken:1982tu,Braaten:1991jj} there are 
analytical studies of collisional parton energy loss, in which highly energetic
partons lose energy and randomise momentum in consecutive partonic $2\to 2$ 
scatterings with partonic components of the medium.  These studies do not
assume particular kinematical constraints.  They are strongly limited, however, 
in that they do not include medium-induced parton emission processes. 

A complementary mechanism, based on analytical studies of radiative parton 
energy loss through modifications of the emission pattern of secondary partons,
was proposed in refs.~\cite{Baier:1996kr,Zakharov:1997uu,Wiedemann:2000za,
Gyulassy:2000er,Wang:2001ifa,Arnold:2002ja}. 
For technical reasons, the emissions in this approach have been limited to 
close-to-eikonal kinematics, neglecting recoil effects.  Also, typically the 
emission of only one gluon from a single highly energetic parton is considered.
Adjusted with a number of phenomenologically motivated requirements, the
simple picture introduced in the publications above has been employed in
model studies in the past years~\cite{Chen:2011vt,Horowitz:2011cv,Majumder:2011uk,Zakharov:2011dq}.
Recent analytical studies improve on some approximations by considering 
gluon emission from eikonal antennae~\cite{
  MehtarTani:2012cy,CasalderreySolana:2011rz}, and by providing a
  differential formulation of the colour flow in the multiple scattering process~\cite{Beraudo:2012bq}.  

In addition, radiative parton energy loss calculations illustrated explicitly 
that by scattering on several spatially separated partons, the distribution 
of emitted gluons is modified by destructive interference effects.  In the 
eikonal limit, these interference patterns are understood in terms of the 
non-Abelian Landau--\-Pomeranchuk--\-Migdal (LPM) effect.  This effect can be 
implemented in a probabilistic formulation based on formation time constraints,
and lead to the first simulations taking this effect consistently into
account~\cite{Zapp:2011ya}.  

By now, however, it is widely accepted that analytical studies of radiative 
parton energy loss provide only very limited constraints on phenomenological 
models implemented in simulation codes~\cite{Armesto:2011ht},  mainly because 
they are restricted to a small part of the relevant kinematics and because 
they only insufficiently constrain how medium-effects modify the vacuum 
structure of the parton shower.  There have been strong efforts in recent 
years to overcome these deficiencies in Monte Carlo formulations of 
medium-modified parton showers~\cite{Deng:2010mv,Lokhtin:2008xi,Zapp:2008gi,
Armesto:2009fj,Armesto:2009ab,Renk:2009nz,Schenke:2009gb,ColemanSmith:2012um}. 
Many of these models use parton formation times to specify destructive 
interference in parton emissions and the spatio-temporal embedding of the 
parton shower in the medium.  This is closely related to the procedure that 
we propose in the present work.  On the other hand, the modelling of 
medium--\-induced gluon emissions typically makes use of analytical results 
obtained in the eikonal limit. Thus, although the Monte Carlo models can 
improve on aspects like implementing exact energy-momentum conservation, it 
should be stressed that they still suffer from the numerically large systematic
uncertainties and conceptual limitations related to using the analytical 
results outside their strict region of validity.  This is why we believe it is
important not only to formulate models with tractable assumptions but also
to supplement them with a well-defined way to quantify related uncertainties.

Finally, analytical studies of parton energy loss in strongly coupled 
non-Abelian plasmas with known gravity duals should be mentioned, for a recent
review cf.~\cite{CasalderreySolana:2011us}.  
This approach is qualitatively different from the one proposed here and we do 
not discuss it further in the present work.

\smallskip

In this paper, a conceptually new approach to jet quenching is formulated,
which consistently includes both collisional and radiative parton energy 
loss mechanisms.  It is based on known properties of perturbative QCD in 
non-eikonal kinematics and their continuation to the infrared region.  This 
formulation consistently treats the scattering of an energetic parton off 
a medium constituent in perturbative QCD using standard techniques.  The 
resulting interaction is described at leading order through standard 
$2\to 2$ QCD matrix elements, which are valid in the full perturbative regime
and can easily be extended into the non-perturbative region with some simple, 
predictive parametrisations.  Further, following the standard description of 
all hard scattering events at colliders, large logarithms associated to 
collinear singularities occurring in matrix elements for subsequent real 
emissions are resummed by the parton shower and thus dealt with in a 
probabilistic language.  This picture provides a well understood and 
systematically improvable approximation to extra gluon emissions.  

In this way both collisional and radiative energy loss are generated 
consistently as part of a more general description of scattering processes in 
perturbative QCD.  This approach includes in a natural way also radiation off 
the scattering centre and generates configurations with any number of 
additional partons.   So-called flavour conversion processes that transfer 
essentially the entire energy of a parton to a different parton are also 
automatically included and the framework can easily accommodate heavy quark 
flavours. 

While this approach offers a general, flexible and largely well controlled 
framework for describing jet evolution in the presence of a dense, strongly 
interacting medium, it is not free of assumptions:

\begin{enumerate}
\item 
  \textbf{The medium as seen by the jet is a collection of partons with a 
    certain distribution in phase space.} 
  For scattering processes with a very high scale (resolution) this is 
  certainly a reasonable assumption. Hard processes happen on very short time 
  scales, which are much shorter than any interaction among the partons 
  forming the medium.  For such processes, the partonic constituents of the medium
  are resolved by the jet, the scattering partner is thus a 
  quasi-free parton. This essentially is the same argument that is employed, 
  for example, in deeply virtual scattering and corresponding factorisation 
  theorems.  The actual assumption here is that this 
  \textbf{factorisation is valid for all the interactions of the jet in the 
    medium}. 
  The actual phase space distribution of scattering partners is irrelevant 
  for the qualitative picture underlying the formulation of jet-medium 
  interactions, but of course quantitatively impact on the results.
\item 
  \textbf{By continuing the pQCD matrix elements into the infra-red region 
    the dominant effect of soft scatterings can be included.} 
  Since it is impossible to restrict the framework in a meaningful way to 
  include perturbatively hard scatterings in the medium only, some modelling 
  of soft scatterings is unavoidable.  While clearly other approaches are 
  conceivable and may be equally legitimate, a suitable infra-red continued 
  version of perturbation theory could be considered as a minimal model, in
  which the exact form of the infra-red continuation and the exact choice of 
  the regulator are a source of uncertainties.  These uncertainties, however,
  can be studied systematically and, with sufficient data, most likely
  be further constrained.  It should be noted, though, that in particular the 
  total scattering cross section in the medium depends on these choices.
\item 
  \textbf{The interplay between different sources of radiation is governed 
    by the formation times of the emissions.}  
  As it is clearly unphysical to assume that additional emissions happen 
  instantaneously, there will be situations where several radiation processes 
  occur at the same time.  In the framework presented here, each emission is 
  ascribed a formation time and for two competing emissions the one with the
  shorter formation time is accepted, while the other will not be realised.
  However plausible this seems, this nevertheless goes beyond standard parton 
  showers, which have no notion of space-time evolution.  Furthermore, there 
  are only parametric estimates of the formation time.  The resulting freedom 
  of an exact choice will yield different admixtures of different processes
  in the same situations.  This, again, gives rise to some uncertainties,
  which, however, are easily quantifiable.
\item \textbf{The physical picture behind the LPM-effect derived from results 
  in the eikonal limit is valid also in non-eikonal kinematics.} 
  In calculations carried out in the eikonal limit it was found that there 
  is a destructive interference between gluon emissions induced by subsequent 
  scattering processes when the distance between the scattering centres is 
  shorter than the gluon formation time.  This is the non-Abelian analogue of 
  the Landau-Pomeranchuk-Migdal (LPM) effect.  A probabilistic interpretation 
  of this phenomenon based on a physical picture was derived and validated 
  in the eikonal limit.  The validity of this picture also in non-eikonal 
  kinematics is only assumed.  A straightforward generalisation of the 
  algorithm can be used to include the LPM-effect in the framework presented 
  here.
\end{enumerate}

\smallskip

In the remainder of this publication, the construction paradigms above will
be further worked out.  To put the different ingredients into perspective, to 
further develop the language employed, and to define our notation, however,
the reader is first reminded of a few concepts that will be used throughout,
cf.~\SecRef{Sec::PS_in_medium}.  
Details of the implementation relevant for a better understanding of our model 
are presented in \SecRef{Sec::Implementation}.
In \SecRef{Sec::Results} we confront this new model with a broad range of 
data, starting from those vacuum observables in $e^-e^+$ and $pp$ collisions,
which fix the parton shower and hadronisation parameters and thus leaves us
with a minimal set of relevant parameters.  By comparing to a number of data
from \textsc{RHIC} and \textsc{LHC}, which show
jet quenching effects, we highlight the versatility of our model.  We also
make a non-trivial, testable prediction there.
Finally, we summarise with some concluding remarks in \SecRef{Sec::Conc}.

\section{Parton shower in a medium}
\label{Sec::PS_in_medium}
In this section, the ingredients necessary to construct an in-medium parton
shower following the paradigms outlined above will be discussed.  In 
\SecRef{Sec::PSReminder} the parton shower picture in the vacuum is briefly 
summarised.  The rescattering of a hard parton in the medium is discussed in 
\SecRef{Sec::ScatInMed}, and in \SecRef{Sec::LPMReminder} the probabilistic 
interpretation of the LPM-effect in the eikonal limit is recapitulated.
Details of its generalisation to the full phase space are given in 
\SecRef{Sec::JewelLPM}.

\subsection{Reminder: parton showers}
\label{Sec::PSReminder}

In the collinear limit, QCD real emission matrix elements and their phase space
factorise such that the differential cross section for any process with an
additional parton in the final state, $\d \sigma_{n+1}$, can be expressed by the
cross section for the production of an $n$-parton final state modified
by process-independent terms:
\begin{equation}
\label{Eq::xsecfac}
 \d \sigma_{n+1} = 
\sigma_n \frac{\d t\d z}{t}\frac{\alphas(\mu^2)}{2 \pi}\hat P_{ba}(z) \,.
\end{equation}
Here $z$ is the energy or light-cone momentum fraction taken by the outgoing
parton, and the splitting kernel $\hat P_{ba}(z)$ typically is given by the 
(unregularised) Altarelli-Parisi splitting function or kernels that reduce to
it in the collinear limit.  The parameter 
$t \approx \kt^2 \approx Q^2 \approx \theta^2$ is a measure for the hardness 
of the additional emission, to leading logarithmic accuracy this can be the 
square of the transverse momentum $\kt$, the virtual mass $Q$ or the emission 
angle $\theta$.  The scale $\mu^2$ at which the strong coupling $\alphas$ is
evaluated is usually given by $\mu^2 = \kt^2$.  While
Eq.~(\ref{Eq::xsecfac}) captures {\bf all leading} collinear and soft-collinear
logarithms irrespective of the precise definition of $t$, the choice 
$t = \theta^2$\footnote{
  In dipole-like showers, also the choice $t=\kt^2$ appears to allow for
  a resummation of such next-to leading logarithms~\cite{
    Gustafson:1987rq}.
} 
and the identification $\mu^2 = \kt^2$ ensure that also the next-to leading 
logarithmic contributions are properly taken into account.

Closer inspection of Eq.~(\ref{Eq::xsecfac}) reveals that the cross section 
diverges for $t \to 0$, as expected.  However, emissions at low scales $t$
of the order of a few $\Lambda_\text{QCD}$ will typically not yield any 
resolvable parton, instead partons radiated at such scales will most likely  
end up in the same hadron as their emitter.  This justifies the introduction 
of an infrared cut-off $\tcut$, usually chosen such that the transverse 
momentum is cut at $\kt^2(\tcut) \approx \unit[1]{GeV^2}$.  This cut-off of 
course also implies a cut-off in $z$ thus shielding the divergent structures 
inside the splitting kernels $\hat P$.

Iterating Eq.~(\ref{Eq::xsecfac}) for any number $k$ of additional emissions 
with these choices, such that
\begin{equation}
\label{Eq::xsecallem}
\d \sigma_{n+k} = \sigma_n\prod\limits_{j=1}^k
\frac{\d t_j\d z_j}{t_j}\frac{\alphas(k_{\perp,j}^2)}{2 \pi}\hat P_j(z_j)
\Theta(t_{j-1}-t_j)\,,
\end{equation}
thus resums, to all orders, terms of the parametric form 
$\alphas^l [L^{2l}+L^{2l-1}]$, where $L$ denotes large logarithms of ratios of 
the scales, $t_0/\tcut$.  It should be noted that this accuracy is achieved 
in the limit, where the number of colours is infinite, $N_c\to\infty$.   
Configurations that are of sub-leading colour are typically suppressed with 
$1/N_c^2$ and come at order $\alphas^l L^{2l-1}$ or below.  

The picture developed so far treats emissions as independent from each other 
(interferences between subsequent emissions are sub-leading) and independent 
from the hard process, apart from the strict ordering  $t_\text{h} = 
t_0>t_1>t_2>\dots$ indicated by the $\Theta$-function in 
Eq.~(\ref{Eq::xsecallem}) and apart from the exact definition of the scale  
$t_\text{h} = t_0$, which indeed depends on the process.  However, it is worth 
noting that the choices of $t\approx \theta^2$ and $t\approx \kt^2$ emerge as 
better suited when inspecting the radiation pattern of two subsequent 
emissions since they better capture effects related to the emergence of 
destructive interference in non-angular-ordered emissions.

The picture above can be cast into a probabilistic form, optimally suited for
detailed simulation by introducing the Sudakov form factor,
\begin{equation}
\label{Eq::sudakov}
 \mathcal{S}_a(t_\text{h},\tcut) =
\exp\left\{-
\int\limits_{\tcut}^{t_\text{h}}\frac{\d t}{t}
\int\limits_{z_\text{min}}^{z_\text{max}}\d z\, \sum_b
\frac{\alphas(\kt^2)}{2\pi}\hat 
P_{ba}(z)
\right\} \,.
\end{equation}
It can be interpreted as the probability that a parton $a$ emits no resolvable 
radiation between the starting scale $t_\text{h}$ and the cut-off $\tcut$.
The independence of the emissions guarantees that the resulting simulation,
the parton shower, can be represented as a Markov chain, i.e.\ the probability 
for any further emission depends only on the current state of the parton 
ensemble and not on the history leading to it. For the generation of radiation 
off final state particles the parton shower starts at the scale of the hard 
core process, each further emissions has to be softer than the previous 
until the infrared cut-off is reached. 

The fact that the result of the evolution is fixed by the hard process at
the higher and by the known initial state, the incident hadron, at the lower 
scale complicates the generation of emissions off initial state particles.  
It would thus be very inefficient to evolve from the soft, hadronic
scale up to the hard scale and discard all histories that do not match the 
hard process.  Instead, initial state partons are evolved backwards starting 
from the hard process with the parton density at each evolution step 
constrained by the parton distribution functions (PDFs).  This modifies the 
Sudakov form factor such that
\begin{equation}
\label{Eq::ISsudakov}
 \mathcal{S}_a^\text{(IS)}(t_\text{h},\tcut,x) = 
\frac{f(x,\tcut)}{f(x,t_\text{h})} 
 \mathcal{S}_a(t_\text{h},\tcut) \,.
\end{equation}
Written in such a form, with suitable Sudakov form factors, the parton shower
correctly resums all leading and dominant sub-leading logarithms.  Various 
methods to further improve the accuracy of the parton shower away from the 
soft and collinear limits of secondary parton radiation exist. However, since 
here we are mainly concerned with the properties of jets, it is sufficient to 
stress that the parton shower, due to its logarithmic accuracy, correctly 
describes the bulk of all QCD events, and in particular the properties of 
jets, which are characterised by multiple soft and collinear emissions.  

\subsection{Scattering in a medium and parton showers}
\label{Sec::ScatInMed}

To analyse how medium effects enter the picture, consider a single scattering 
of an energetic projectile off a parton in the medium at a time that is late 
enough such that the parton shower evolution of the initial hard scattering 
process is finished and all emerging final state partons are on their mass 
shell.  This process fundamentally is not different from the hard initial 
process that produced the hard parton in the first place.  Only the 
centre-\-of-\-mass energy and thus the scale of the scattering are, on 
average, diminished and since there is no natural infrared cut-off as in the 
case of the initial hard scatter (the criterion that the jets should have at 
least a certain $\pt$) one has to specify how the infrared region is to be 
treated.  In \textsc{Jewel} an infrared continuation of the matrix elements is 
introduced, which will be discussed in section~\ref{Sec::Implementation}. 
Then, just as in the case of the initial hard jet production process, leading 
order $2\to 2$ matrix elements supplemented with parton showers provide a 
powerful approximation to higher--\-order real emission matrix elements.  

The only difference with respect to the case already discussed in 
\SecRef{Sec::PSReminder} relevant for the implementation of such a picture
is that the incident partons in such a secondary scatter are not constituents 
of an incoming nucleon. 
Therefore, using nucleon PDFs in the initial state parton shower is 
inappropriate and instead 'partonic PDFs' must be invoked to properly account
for the fact that in the backwards evolution of the initial state shower
the incident particles are known.  These parton PDFs are constructed to 
describe the densities of partons in partons, by taking into account possible 
radiation above the resolution scale $\tcut$.   At and below $\tcut$ no 
radiation is possible and the parton has no associated parton density.  The 
partonic PDFs are thus computed by integrating the DGLAP equations with the 
boundary conditions 
\begin{equation}
 f_i^j(x,\tcut) = \left\{ 
\begin{array}{lll} \delta(1-x) & \quad ; & i=j \\
 	0 & \quad ; & i\neq j 
\end{array} \right. \, .
\end{equation}

\medskip

In this way $2\to 2$ configurations without additional radiation, that would 
be classified as elastic processes, and configurations with 3 and more final 
state particles, that would be regarded as inelastic processes or scattering 
with medium-induced Bremsstrahlung are generated at the same time, with no
additional modelling bias.  It should also be noted that matrix elements for 
different multiplicities of final state partons cannot be naively combined 
due to double counting issues~\cite{Catani:2001cc}, which are typically 
rarely addressed in the phenomenology of jet quenching effects.  
More importantly, though, these processes all come with the (leading log) 
correct relative rates (the total scattering rate is defined by the LO cross 
section as the parton shower is unitary).  In principle, hard emissions could 
also be corrected to the full matrix elements using merging prescriptions 
such as \cite{Catani:2001cc,Mangano:2006rw}, but for the purpose of this 
study the accuracy of the parton shower is sufficient.  A further advantage 
of this description is that it is not restricted to eikonal or close-to-eikonal 
kinematics (coherence issues will be discussed in \SecRef{Sec::JewelLPM}). 

\medskip

A complication arises when a rescattering occurs before the parton shower 
evolution of the previous scattering (initial or rescattering in the medium) 
has terminated. The interplay of such different sources of radiation in 
\textsc{Jewel} is governed by formation times. The formation time associated 
with an emission parametrically is of the form 
\begin{equation}
\tau = \frac{E}{t} \,,
\end{equation}
where $E$ is the emitting parton's energy\footnote{
  This is parametrically equivalent to the form $\tau = 2 \omega/\kt^2$, 
  familiar from analytical calculations of coherent Bremsstrahlung in the 
  eikonal limit.}.  
In the case of rescattering during the parton shower evolution, it is the 
emission with the shorter formation time that is formed, while the other
emission will be discarded.  This means in turn that sufficiently hard 
rescatterings reset the parton shower and restart it at a scale given by
their kinematics, while soft rescatterings cannot initiate further radiation.  

To further illustrate these points, consider a situation where a parton in 
the parton shower (it does not matter which scattering initiated this parton 
shower) is going to split at a scale $t_1$ on a timescale $\tau_1$, while a 
further scattering with scale $t_2^\text{(s)}$ just happens before that.  
Then a hypothetical parton shower with starting 
scale $t_2^\text{(s)}$ is initiated.  If the hypothetical parton shower does not 
radiate, then the original splitting at $t_1$ proceeds as foreseen. If, on the 
other hand, the hypothetical parton shower would initiate radiation at a scale 
$t_2$ with corresponding formation time $\tau_2$, then there are two possible 
outcomes:  If $\tau_2 > \tau_1$ the hypothetical shower is discarded and only
the splitting at $t_1$ happens.  In the opposite case $\tau_2 < \tau_1$ the 
original splitting at $t_1$ is discarded, the hypothetical shower becomes real 
and replaces the original one.  The splitting at $\tau_2$ takes place and any 
further radiation from the new parton shower proceeds, provided it is not 
disturbed by further rescattering.  This formation time prescription is on 
average equivalent to the statement that only rescatterings which are harder 
than the current parton shower scale can resolve the virtual parton. As a 
consequence, individual emissions cannot be unambiguously associated with a 
certain scattering. 

\subsection{Reminder: probabilistic interpretation of the LPM effect}
\label{Sec::LPMReminder}

Analytical calculations of medium induced gluon radiation in the eikonal limit 
found that the non-Abelian analogue of the Landau-Pomeranchuk-Migdal (LPM) 
effect plays an important role.  This is a destructive interference between 
emissions initiated by subsequent scattering processes that occurs when the 
formation times of the individual emissions overlap.  In~\cite{Zapp:2011ya} a 
probabilistic formulation suitable for Monte Carlo implementation was derived, 
the findings relevant for this discussion will be summarised here. 

In the eikonal limit, the radiating parton has asymptotically high energy $E$. 
The medium constituents then appear as static scattering centres and the 
momentum transfer between the projectile and the scattering centres is purely 
transverse.  The original BDMPS approach~\cite{Baier:1996kr} and similar other 
approaches~\cite{Zakharov:1997uu,Wiedemann:2000za,
Gyulassy:2000er,Wang:2001ifa,Arnold:2002ja} 
operate in a kinematical regime where $E\gg \omega \gg \kt, \qt$, where $\kt$ 
denotes the transverse momentum of the gluon and $\qt$ the momentum transfer 
from the scattering centre \footnote{The BDMPS-Z formula for parton energy 
loss also holds in the kinematical region $\omega \gg E-\omega \gg \kt, \qt$, 
but an explicit derivation without strong energy ordering is not known. We 
thank P. Arnold for this comment.}.  
Gluon radiation is described by a Gunion-Bertsch 
cross section of the form
\begin{eqnarray}
 \frac{\d \sigma^\text{(GB)}}{\d \mathbf{k}_\perp \d \mathbf{q}_\perp} \propto 
\frac{\mathbf{q}_\perp^2}{\mathbf{k}_\perp^2 (\mathbf{k}_\perp - 
\mathbf{q}_\perp)^2} \,.
\end{eqnarray}
For the sake of deriving and validating a probabilistic interpretation of 
gluon radiation and in particular the LPM-effect it is entirely sufficient to 
keep only the singular part (i.e.\ set $\mathbf{k}_\perp = \mathbf{q}_\perp$) 
and ascribe all elastic rescatterings to the gluon (as only the transverse 
momentum of the gluon relative to the projectile is relevant).

A radiated gluon decoheres from the projectile when it has accumulated a 
relative phase
\begin{equation}
 \varphi = \frac{\kt^2}{2 \omega} \Delta L
\end{equation}
of order unity. Here, $\Delta L$ is the distance it travelled. From this 
condition the gluon formation time $\tau = 2\omega/\kt^2$ can be read off. 

Inspection of the calculations reveals that the scatterings occurring during 
the gluon formation time act coherently to radiate the gluon. In the gluon 
radiation term only the vector sum 
$\mathbf{Q}_\perp = \sum_i \mathbf{q}_{\perp,i}$ 
of the individual momentum transfers appears. 

The probabilistic picture is that whenever an additional scattering occurs 
during the formation time, the respective momentum transfer is coherently 
added. There are two ways how this can be achieved in practice:  The first
one is to update the gluon phase accumulated up to the last scattering centre 
based on the gluon momentum before the last momentum transfer.  Then the 
total $\mathbf{Q}_\perp$ including the last scattering is computed and the 
gluon $\kt$ is changed accordingly.  The new formation time is determined 
taking the already existing phase into account.  The other option takes the 
multiple scattering during the formation time less literally.  In this case, 
whenever a further momentum transfer is added, the new $\mathbf{Q}_\perp$ 
and $\kt$ are computed and the gluon formation time given by the updated 
$\kt$ is calculated.  If the last scattering still is within the updated 
formation time limit (counted from the first scattering) it is accepted, 
otherwise it is rejected as it leads to an inconsistent configuration.  
Finally, in both cases, when there are no more coherent momentum transfers 
the gluon emission is accepted with probability $1/N_\text{scat}$, where 
$N_\text{scat}$ is the number of coherent scatterings.

\section{Details of the implementation}
\label{Sec::Implementation}

\subsection{Event generation}
In our model, event generation proceeds as follows: 
\begin{enumerate}
\item The hard matrix elements initially producing the di-jets and the 
  corresponding initial state parton shower are generated by 
  \textsc{Pythia}~6.4~\cite{Sjostrand:2006za} running with the 
  virtuality--ordered parton shower to provide the best fit to the 
  \textsc{Jewel} simulation. There are no multi--parton interactions and the 
  \textsc{Cteq6l1}~\cite{Pumplin:2002vw} PDFs as provided by 
  \textsc{Lhapdf}~\cite{Whalley:2005nh} are used; simulating nuclear collisions 
  the nuclear modification of \textsc{Eps09lo}~\cite{Eskola:2009uj} is 
  employed. 
\item Then \textsc{Jewel} takes over and selects the impact parameter of the 
  event according to the geometrical cross section and the transverse position 
  of the hard scattering according to the density of binary nucleon-nucleon 
  interactions in the transverse plane. It then generates the final state 
  parton showers including interactions in the medium.  This includes
  rescatterings, emission of secondary partons, an implementation of the
  LPM effects and so on.  These details will all be discussed below.
\item Finally, the event is handed back to \textsc{Pythia} for hadronisation. 
\end{enumerate}
Note that in the absence of a medium the procedure is the same and the 
\textsc{Jewel} parton shower becomes an ordinary vacuum parton shower.  This 
option is used for the validation of the \textsc{Jewel} parton shower, 
for the generation of the p+p baseline, and for the fixing of shower and
hadronisation parameters.

In the following we will highlight aspects of phase 2 of the sketch above.

\subsection{In--medium matrix elements and their infrared continuation}
The interaction of the parton shower with the partonic constituents of the 
medium depends on properties of the medium. We specify the cross sections for 
$2 \to 2$ processes as
\begin{equation}
\label{eq:crosssection}
\sigma_i(E,T) = 
\int\limits_0^{|\hat t|_{\text{max}}(E,T)}\!\!\!\!\! \d |\hat t|\!\! 
 \int\limits_{x_{\text{min}}(|\hat t|)}^{x_{\text{max}}(|\hat t|)} \!\!\!\!\! \d x
\sum_{j \in \{\text{q,\=q,g}\}} \!
f_j^i(x, \hat t) \frac{\d \hat \sigma_j}{\d \hat t}(x\hat s,|\hat t|) \,,
\end{equation}
where the PDF takes into account possible initial state radiation off the 
energetic projectile. Note that here, implicitly, we neglect a similar 
evolution experienced by the target, i.\,e.\ initial state radiation off the 
medium parton.  This choice, although inconsistent at first glance, is 
motivated by the fact that these partons have only very small energy in
the laboratory frame, which translates into them being virtually unable to
emit any resolvable parton.  

The maximum momentum transfer $|\hat t|_\text{max}$ is determined by the initial
kinematics of the scattering. Neglecting the scattering centre's momentum 
$|\hat t|_\text{max} = 2 m_\text{s}(T) [E_\text{p} - m_\text{p}]$, 
where $m_\text{s}(T)$ stands for the (temperature dependent) 
scattering centre's mass and $E_\text{p}$ and $m_\text{p}$ are the projectile 
parton's energy and (virtual) mass, respectively. The boundaries on the $x$-integral are 
obtained from the requirement that $\kt^2 \ge Q_0^2/4$ and are given by 
$x_{\text{min}}(|\hat t|) = Q_0/(4|\hat t|)$ and $x_{\text{max}}(|\hat t|) = 1 
- Q_0/(4|\hat t|)$. For the partonic cross section we keep leading terms only, 
but we regularise them with a Debye mass $\mu_\text{D} \approx 3 T$. They 
therefore read
\begin{equation}
 \frac{\d \hat \sigma}{\d \hat t}(\hat s,|\hat t|) = C_{\text{R}}
\frac{\pi}{\hat s^2} \alphas^2(|\hat t| + \mu_{\text{D}}^2)\frac{\hat s^2 + (\hat
s-|\hat t|)^2}{(|\hat t| + \mu_{\text{D}}^2)^2} \longrightarrow C_{\text{R}}
2 \pi \alphas^2(|\hat t| + \mu_{\text{D}}^2)\frac{1}{(|\hat t| +
\mu_{\text{D}}^2)^2} \,.
\label{cross}
\end{equation}

\subsection{Parton shower}
The \textsc{Jewel} parton shower~\cite{Zapp:2008gi} is a virtuality--ordered 
monopole shower, similar to the \textsc{Pythia}~6 virtuality ordered 
shower~\cite{Sjostrand:2006za}, but without applying, ``by hand'' the a 
posteriori angular ordering constraint, when being run in the medium.
Identifying the evolution parameter $t$ with the virtual mass squared $Q^2$ of 
the partons in the shower, the infrared cut-off is denoted by $Q_0^2$, and 
the splitting parameter $z$ denotes the energy splitting.  In this setup, the 
transverse momentum of the splitting is given by $\kt^2 \simeq z (1-z) Q^2$ 
for time-like and $\kt^2 \simeq (1-z) Q^2$ for space-like splittings.  The 
current implementation is limited to massless quarks, the extension to 
massive quarks is straightforward.

Contrary to the way in which the \textsc{Pythia}~6 shower is implemented, the 
kinematics of each splitting is already fully constructed when the splitting 
is generated. This requires determining the daughter virtualities together 
with the splitting of the mother. For parton shower evolution in the vacuum 
this procedure is more involved than the traditional approach of later 
correcting the kinematics of the splitting of the mother, but is advantageous 
for jet evolution in a medium. The reason for this is that in the medium one 
evolves not only in the shower ordering variable but also in time and going 
back to the mother splitting means going back in time, which is very cumbersome.

\smallskip

Since the centre-of-mass energies of scatterings in the medium are 
typically rather small, it is unlikely that more than one initial state 
splitting is initiated in secondary scatters.  In this approximation\footnote{
  This approximation used here can easily be improved numerically, adding
  more emissions, if necessary.} 
the partonic PDFs can be integrated analytically yielding
\begin{align}
f_\text{q}^\text{q}(x,Q^2) = & \mathcal{S}_\text{q}(Q^2,Q_0^2) \delta(1-x)
 + \int\limits_{Q_0^2}^{Q^2}
\! \frac{\d q^2}{q^2}\, \mathcal{S}_{\text{q}}(Q^2,q^2)
\, \frac{\alphas((1-x)q^2)}{2\pi} {\hat P}_{\text{qq}}(x)\,, \\
f_\text{\= q}^\text{q}(x,Q^2) = & 0 \\
f_\text{g}^\text{q}(x,Q^2) = & \int\limits_{Q_0^2}^{Q^2}
\! \frac{\d q^2}{q^2}\, \mathcal{S}_\text{g}(Q^2,q^2)
\, \frac{\alphas((1-x)q^2)}{2\pi} \hat P_\text{gq}(x)\,, \\
f_\text{q}^\text{g}(x,Q^2) = & f_\text{\=
q}^\text{g}(x,Q^2) =
\int\limits_{Q_0^2}^{Q^2}
\! \frac{\d q^2}{q^2}\, \mathcal{S}_\text{q}(Q^2,q^2)
\, \frac{\alphas((1-x)q^2)}{2\pi} \hat P_\text{qg}(x)\,, \\
f_\text{g}^\text{g}(x,Q^2) = &
\mathcal{S}_\text{g}(Q^2,Q_0^2) \delta(1-x) \notag\,, \\
 & \ + 2 \int\limits_{Q_0^2}^{Q^2}
\! \frac{\d q^2}{q^2}\, \mathcal{S}_\text{g}(Q^2,q^2)
\, \frac{\alphas((1-x)q^2)}{2\pi} \hat P_\text{gg}(x) \,.
\end{align}

\subsection{Implementation of the LPM effect in JEWEL}
\label{Sec::JewelLPM}

In \textsc{Jewel} it is assumed that the physical reasoning behind the 
probabilistic formulation of the LPM-effect is valid also outside the eikonal 
limit.  Although the kinematics is very different, the procedure of adding 
scattering processes within the gluon formation time coherently by adding 
their momentum transfers and reweighting the actual emissions proceeds in 
exactly the same way. 

A slight complication arises due to the fact, that -- in contrast to the 
eikonal case -- the formation time is not directly related to the momentum 
transfer but to the scale of the emission, which can be anything between the 
scale $t^\text{(s)}$ set by the momentum transfer and $\tcut$.  The procedure 
for adding a momentum transfer coherently and determining the scale of the 
associated radiation is outlined below. Here, $t_1^\text{(s)}$ denotes the 
scale of scattering before that last momentum transfer is added and 
$t_2^\text{(s)}$ is the scale including the last momentum transfer.  As the 
momentum transfers are added vectorially, the resulting $t_2^\text{(s)}$ may 
be larger or smaller than $t_1^\text{(s)}$.  Before the last scattering, 
$t_1^\text{(s)}$ was the starting scale of a parton shower and (provided 
$t_1^\text{(s)} > t_\text{c}$) the parton shower may or may not foresee a 
gluon radiation at a scale $t_1 < t_1^\text{(s)}$.  Updating the starting 
scale set by the effective (coherent) scatterings to $t_2^\text{(s)}$ may 
increase or decrease the phase space for radiation and the existing emission 
(if there is one) has to be corrected for the corresponding change in phase 
space and emission probability. As a result an emission may have to be 
rejected, it may be assigned a new scale $t_2$, or a new emission may be
enforced where there was none before. All the different cases are 
summarised below.

\begin{itemize}
 \item $t_2^\text{(s)}<\tcut$: \\	
   In this situation no radiation is possible, if there is an existing
   emission it has to be rejected.
 \item \textit{existing radiation}, $t_2^\text{(s)}>t_1^\text{(s)}$: \\
   The probability for an emission is now larger and so the existing
   radiation cannot be rejected. The scale of the radiation, however, may
   have to be redetermined. The chosen $t_1$ for the emission is kept 
   with probability $\mathcal{S}(t_2^\text{(s)},t_1^\text{(s)})$, i.e.\ if 
   there is no radiation between $t_1^\text{(s)}$ and $t_2^\text{(s)}$ it has to 
   be in the old interval $[\tcut,t_1^\text{(s)}]$. In the other case where 
   we have radiation between $t_1^\text{(s)}$ and $t_2^\text{(s)}$ a new scale 
   $t_2$ in this interval is chosen and the old one has to be discarded.
 \item \textit{existing radiation}, $t_1^\text{(s)}>t_2^\text{(s)}>\tcut$: \\
   In this case the probability for radiation has decreased and the
   existing emission has to be rejected with probability
   $[1-\mathcal{S}(t_1^\text{(s)},\tcut)]/[1-\mathcal{S}(t_2^\text{(s)},\tcut)]$.
   If the emission is kept but its scale $t_1$ is larger than $t_2^\text{(s)}$ 
   a new scale $t_2$ in the new interval has to be chosen.
 \item \textit{no existing radiation}, $t_2^\text{(s)}>t_1^\text{(s)}$: \\
   The phase space for emission has increased and with probability
   $1-\mathcal{S}(t_2^\text{(s)},t_1^\text{(s)})$ a new emission with scale $t_2$
   to be chosen between $t_1^\text{(s)}$ and $t_2^\text{(s)}$ is added.
 \item \textit{no existing radiation}, $t_1^\text{(s)}>t_2^\text{(s)}>\tcut$: \\
   In this case nothing happens since the probability for radiation has
   decreased and no emission needs to be reweighted. 
\end{itemize}

The reweighting of the emission with $1/N_\text{scat}$ is straightforward 
and, as discussed in section~\ref{Sec::ScatInMed}, it will only be accepted if 
its final formation time is shorter than that of potential other emissions. 

\subsection{Recoils, colour flows, and hadronisation}
In \textsc{Jewel} only the QCD evolution of jets and their interactions with 
the medium are simulated and not the complete event. The recoiling scattering 
centres can be traced and in principle they could be subjected to further 
interactions. This option is, however, disabled by default as these 
interactions are typically much softer than the jet-medium interactions and 
perturbation theory is not necessarily the correct language to describe them 
(and due to the rapid increase in number of interactions the run-time increases 
dramatically). As a way of estimating the effect of the jet on the medium and 
jet-background correlations the recoils may be kept in the event, removed or 
replaced with uncorrelated scattering centres. 

As in the vacuum parton shower every radiated gluon is a colour neighbour of 
the emitter.  When the recoils are not traced they also don't exchange colour 
with the jet.  When the recoils are kept in the event, they are assumed to be 
neighbours of their scattering partner and are inserted in the existing colour 
string. This obviously constitutes a simplification but avoids another 
complication.  After all splittings and scatterings the event is hadronised 
using \textsc{Pythia}'s implementation of the Lund string fragmentation. For 
the Lund string model to work properly it is essential that the strings are of 
sufficient invariant mass.  Naively trying to track all colour indices of all 
scattering centres without invoking any colour reconnection model indeed 
leads to large numbers of strings with very low invariant mass and the 
breakdown of the hadronisation model.

\subsection{Medium model}

\textsc{Jewel} can in principle be interfaced with any medium model that 
specifies the density and momentum distribution of scattering centres for any 
point in space-time. In order 
to have a simple, highly predictive model capturing all essential features of
the medium without introducing too much additional, potentially intractable
kinematics and, at the same time, allowing full control over the medium, we 
used a variant of the Bjorken model~\cite{Bjorken:1982qr,Zapp:2005kt} for this study.  It 
describes the boost-invariant longitudinal expansion of an ideal 
quark--gluon--gas with three parameters only, namely the initial proper 
time $\taui$ at which the hydrodynamic evolution sets in, the initial 
temperature $\ti = T(\taui)$ and the critical temperature $\tc$ of the 
deconfinement phase transition.  The transverse profile is chosen such that 
the energy density is proportional to the density of wounded nucleons. The 
latter is, as all geometrical aspects, calculated in the framework of a simple 
Glauber model~\cite{Eskola:1988yh}. In 
practice this means that the initial temperature is translated into an initial 
energy density $\epsilon_\text{i} \propto \ti^4$.  The actual energy density 
profile at $\taui$ is then given by
\begin{equation}
\epsilon(x,y,b,\taui) = \epsilon_\text{i} 
  \frac{n_\text{part}(x,y,b)}{\langle n_\text{part} \rangle (b=0)}
  \qquad \text{with} \qquad 
  \langle n_\text{part} \rangle (b=0) \approx \frac{2A}{\pi R_A} \,,
\end{equation}
where $b$ is the impact parameter, $n_\text{part}(x,y,b)$ is the density of 
participating nucleons in the transverse plane and we have for simplicity 
assumed a symmetric $A+A$ collision.  As a consequence the temperature is 
higher in the centre of the overlap region and it decreases with centrality. 
The (proper) time dependence is given by 
\begin{equation}
\epsilon(x,y,b,\tau) = \epsilon(x,y,b,\taui)
         \left(\frac{\tau}{\taui}\right)^{-4/3}
 \quad \text{and} \quad 
 T(x,y,b,\tau) \propto \epsilon^{1/4}(x,y,b,\taui) 
  \left(\frac{\tau}{\taui}\right)^{-1/3} \,.
\end{equation}
The time dependence of the particle density is then given by 
$n(x,y,b,\tau) \propto T^3(x,y,b,\tau)$.
We fix the critical temperature at $\tc = \unit[165]{MeV}$ consistent with 
lattice results and since we are only considering interactions in the 
deconfined phase there are no scatterings when the local temperature has 
dropped below $\tc$.

\section{Results}
\label{Sec::Results}

The analysis of Monte Carlo events and all plots shown here were produced 
with Rivet~\cite{Buckley:2010ar} and the corresponding analysis codes contained
within.  For jets, typically the anti-$\kt$ algorithm of~\cite{Cacciari:2011ma}
has been used throughout.
\subsection{Validation}

\begin{figure}[t]
 \centering
 \includegraphics[width=0.48\textwidth]{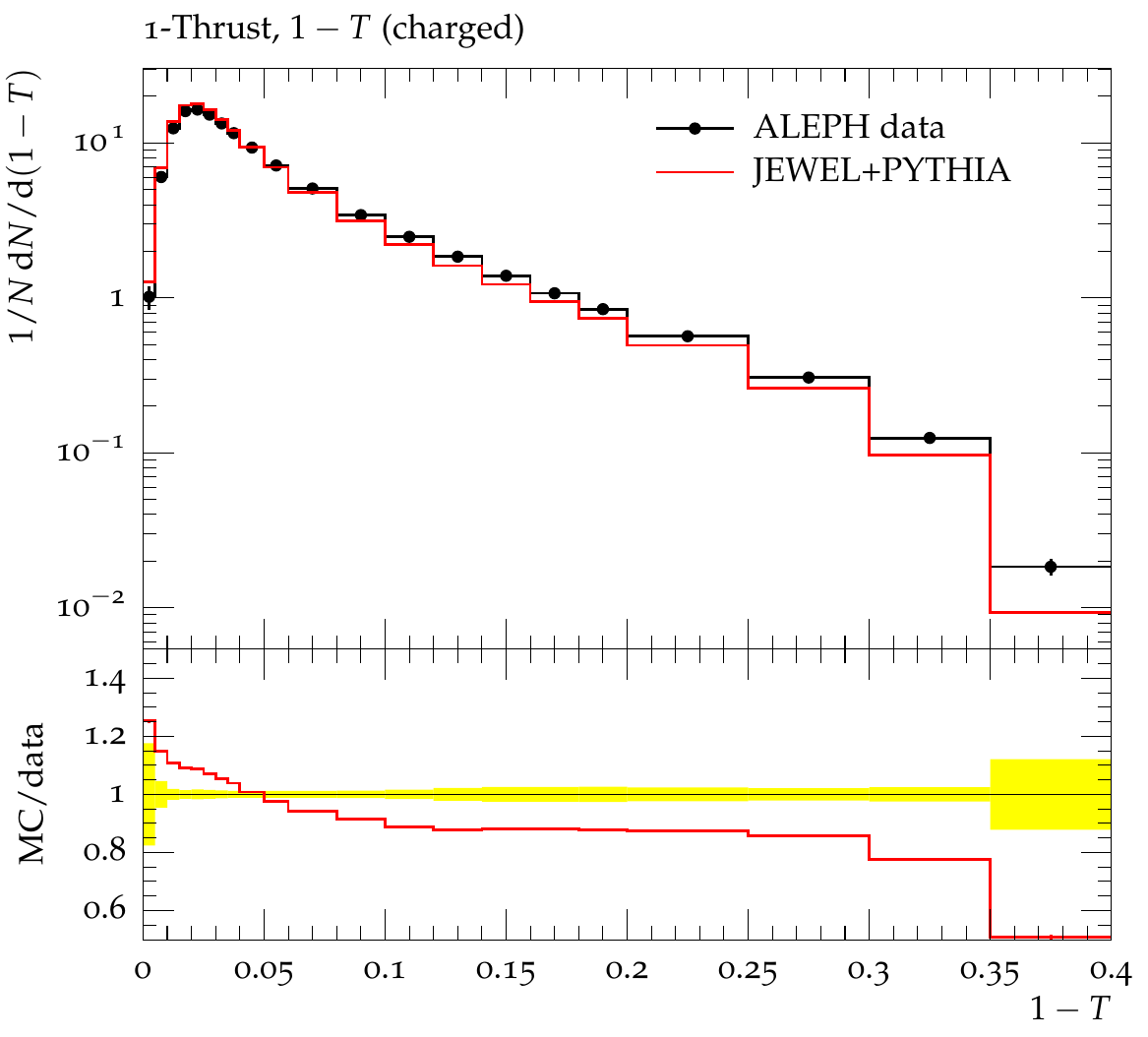}
 \includegraphics[width=0.48\textwidth]{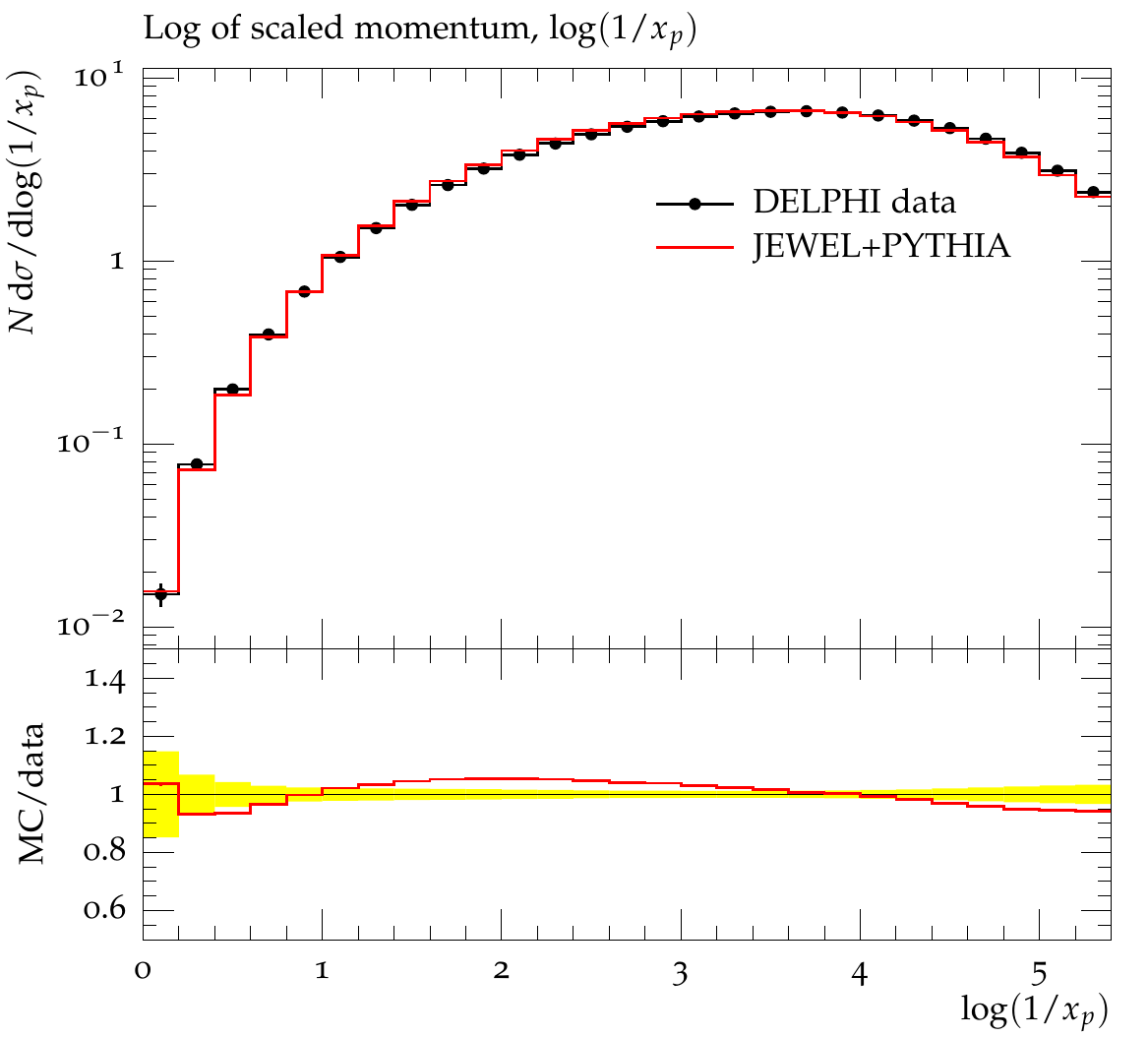}
 \caption{Comparison of \textsc{Jewel+Pythia} results to \textsc{Lep} data. 
   \textbf{LHS}: Thrust distribution measured by 
   \textsc{Aleph}~\cite{Barate:1996fi}.
   \textbf{RHS}: charged particle fragmentation function as a function of the 
   scaled momentum $x_\text{p} = 2 p_\text{h}/\sqrt{s}$ measured by 
   \textsc{Delphi}~\cite{Abreu:1996na}.}
\label{Fig::LEP}
\end{figure}

In the absence of a medium \textsc{Jewel} reduces to a standard (vacuum) parton
shower, which was validated extensively against data from \textsc{LEP} and 
$p+p$ collisions at \textsc{LHC}. 

In all results shown here, the strong coupling is running at one loop with 
$\alphas(m_{\text{Z}})=0.128$, consistent with findings of other leading order 
parton shower calculations.  The infrared cut-off of the parton shower was 
chosen as $Q_0 = \unit[1.5]{GeV}$. As the \textsc{Jewel} parton shower is 
very similar to the \textsc{Pythia} virtuality--\-ordered shower and given 
the very reasonable description of the \textsc{LEP} data, a retuning of the 
hadronisation parameters was considered not to be necessary. 

As examples illustrating the generally very satisfactory performance of the 
parton shower (and hadronisation) at \textsc{LEP} the thrust distribution and 
the charged particle fragmentation function are shown in Fig.~\ref{Fig::LEP}. 

\begin{figure}[t]
 \centering
 \includegraphics[width=0.48\textwidth]{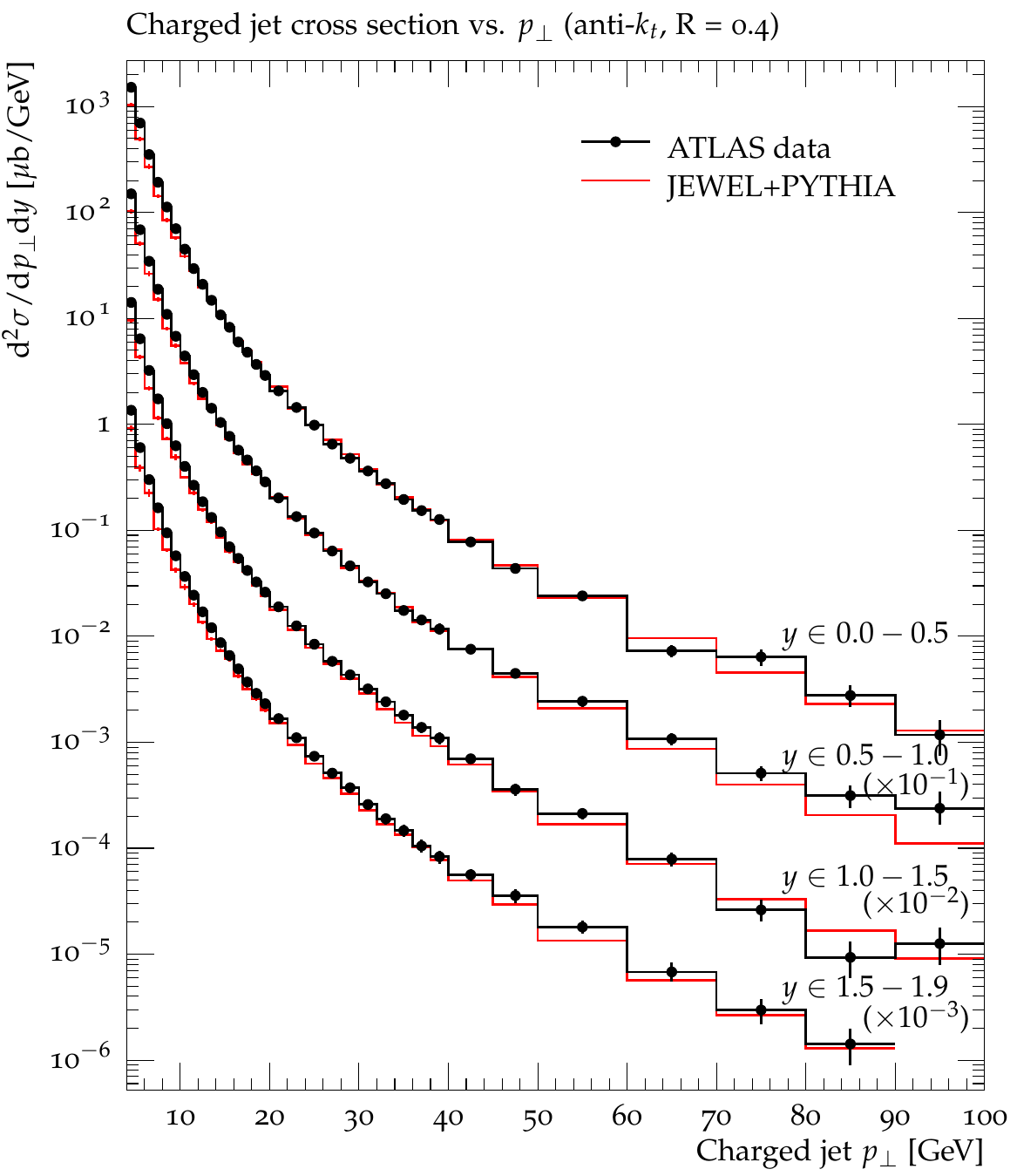}
 \includegraphics[width=0.48\textwidth]{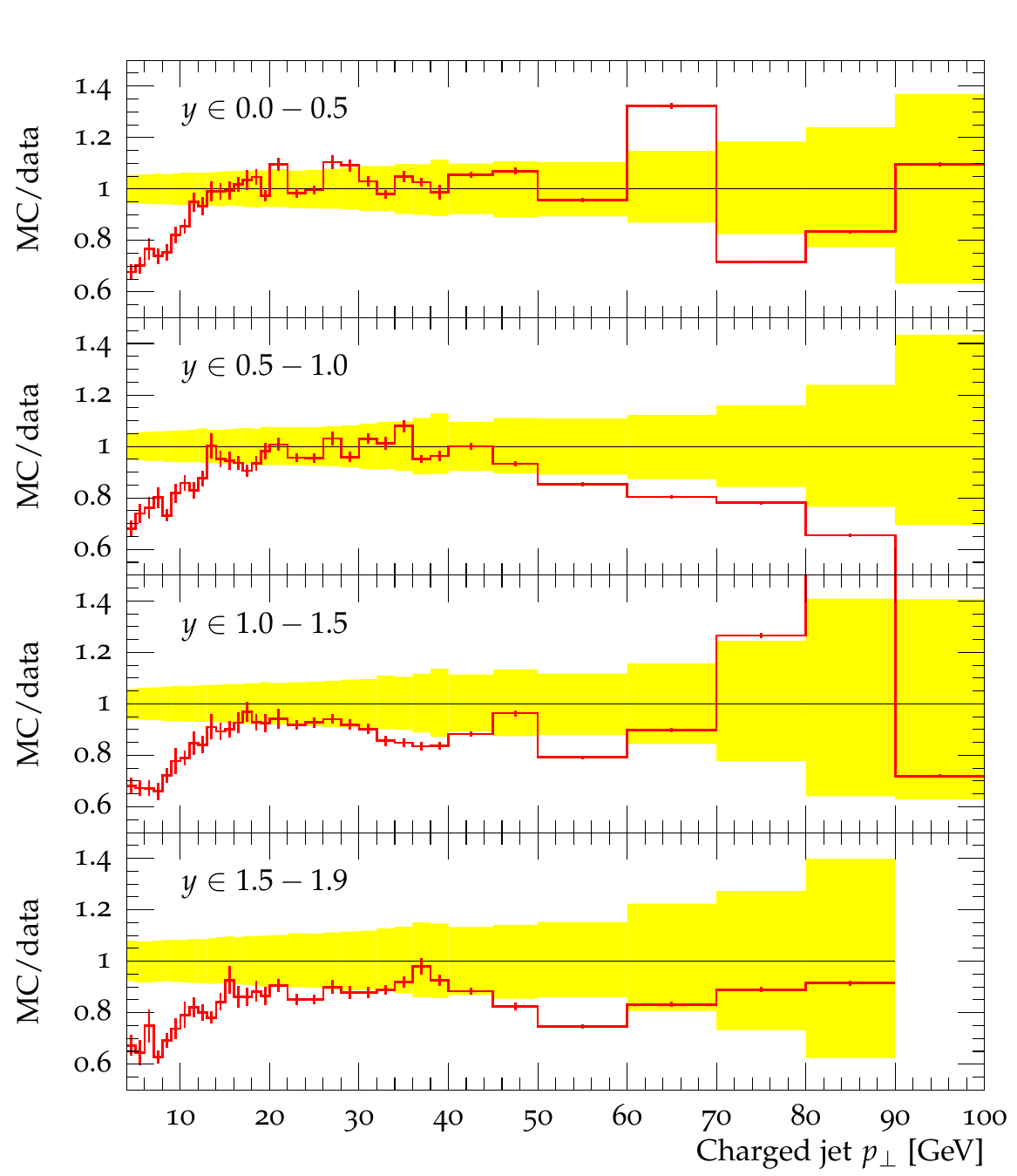}
 \caption{\textsc{Jewel+Pythia} results for the inclusive jet cross section 
   in different rapidity bins compared to \textsc{Atlas} 
   data~\cite{Aad:2011gn} in p+p at $\sqrt{s}=\unit[7]{TeV}$.  Jets 
   are reconstructed using the anti-$\kt$ algorithm with R=0.4 on 
   tracks.}
\label{Fig::ppatLHCjetspec}
\end{figure}

\begin{figure}[t]
 \centering
 \includegraphics[width=0.48\textwidth]{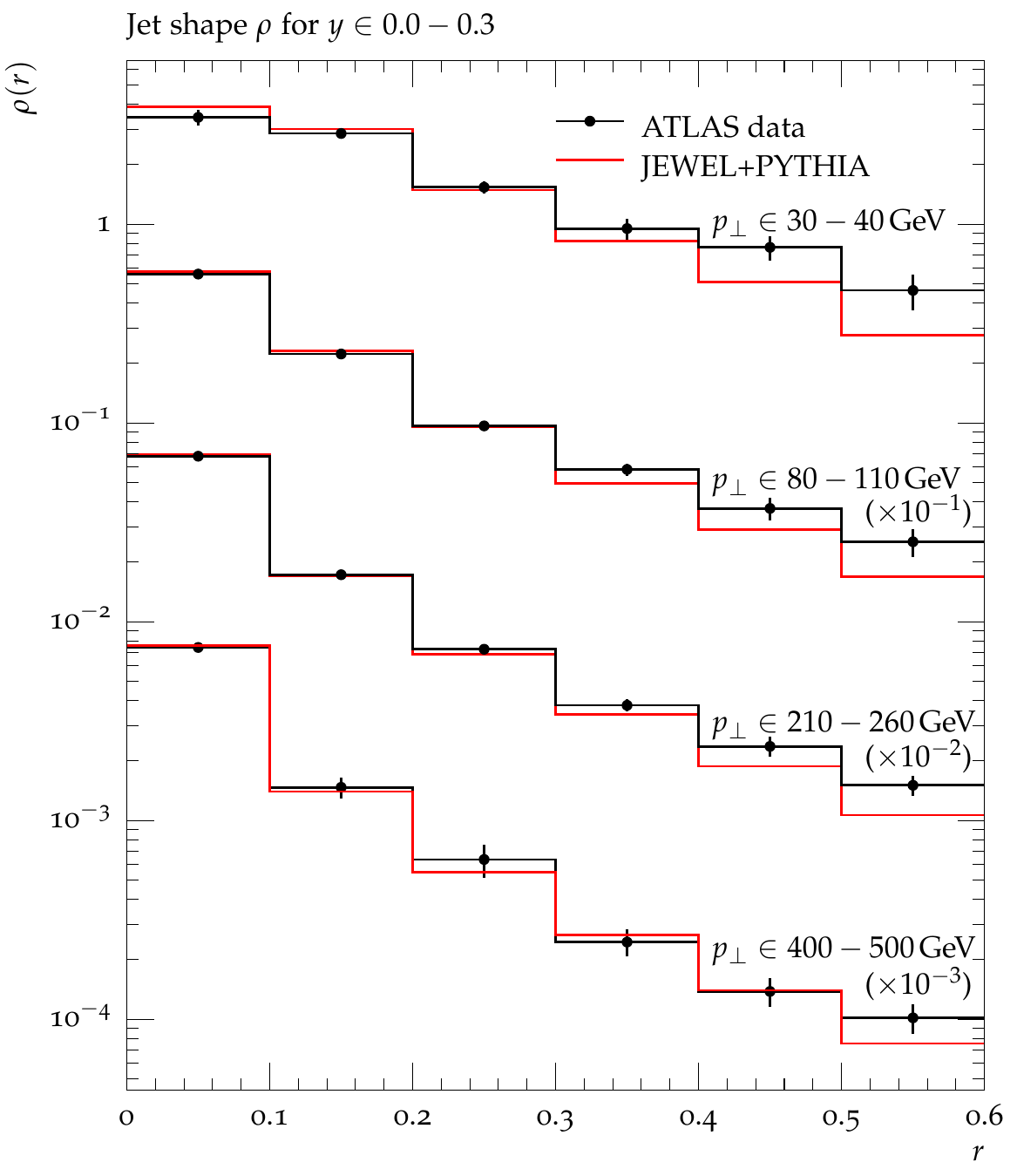}
 \includegraphics[width=0.48\textwidth]{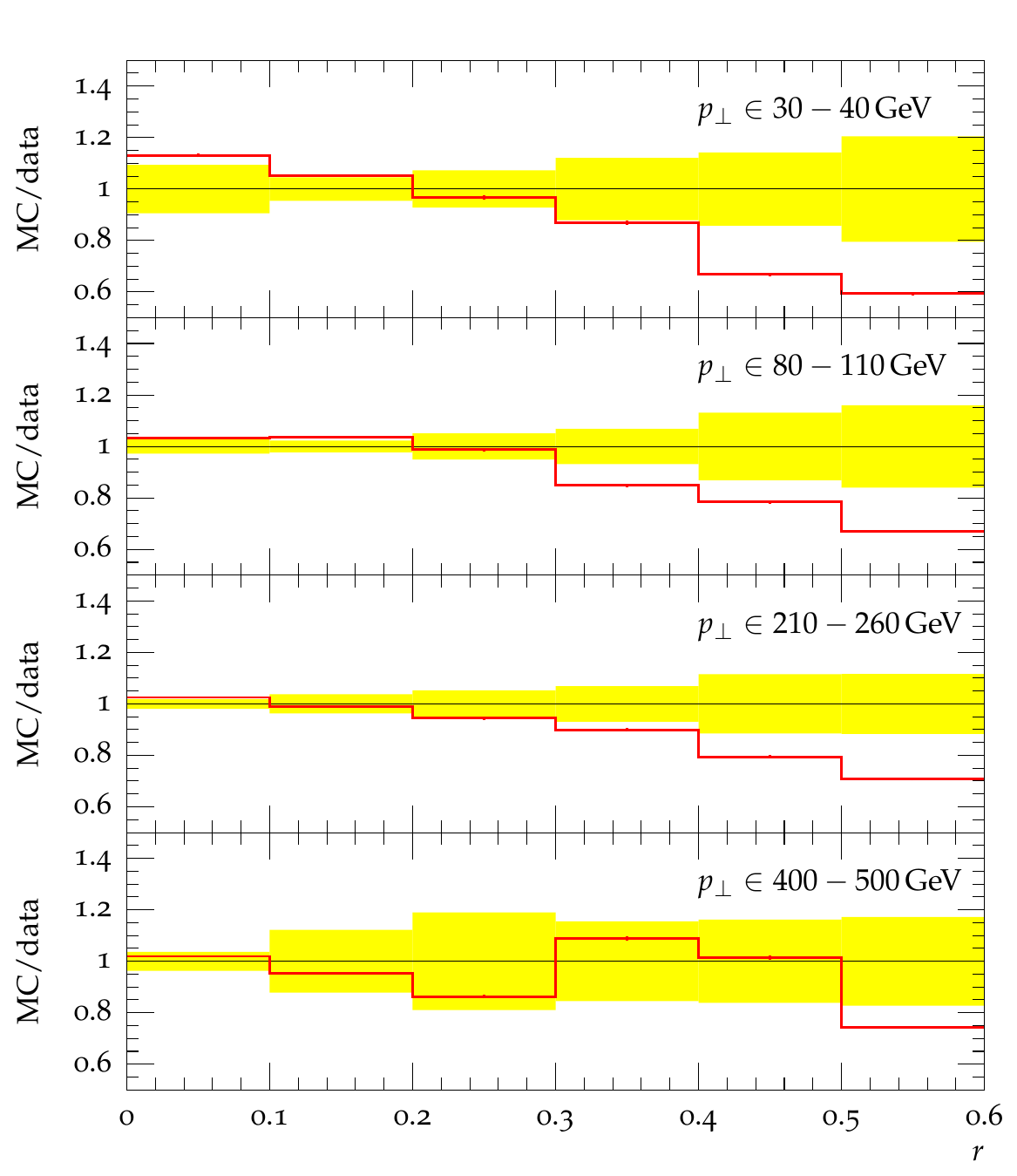}
 \caption{\textsc{Jewel+Pythia} compared to \textsc{Atlas} measurements of 
   the jet shape $\rho(r)$~\cite{Aad:2011kq} in p+p at 
   $\sqrt{s}=\unit[7]{TeV}$.}
\label{Fig::ppatLHCjetshape}
\end{figure}

\begin{figure}[t]
 \centering
 \includegraphics[width=0.48\textwidth]{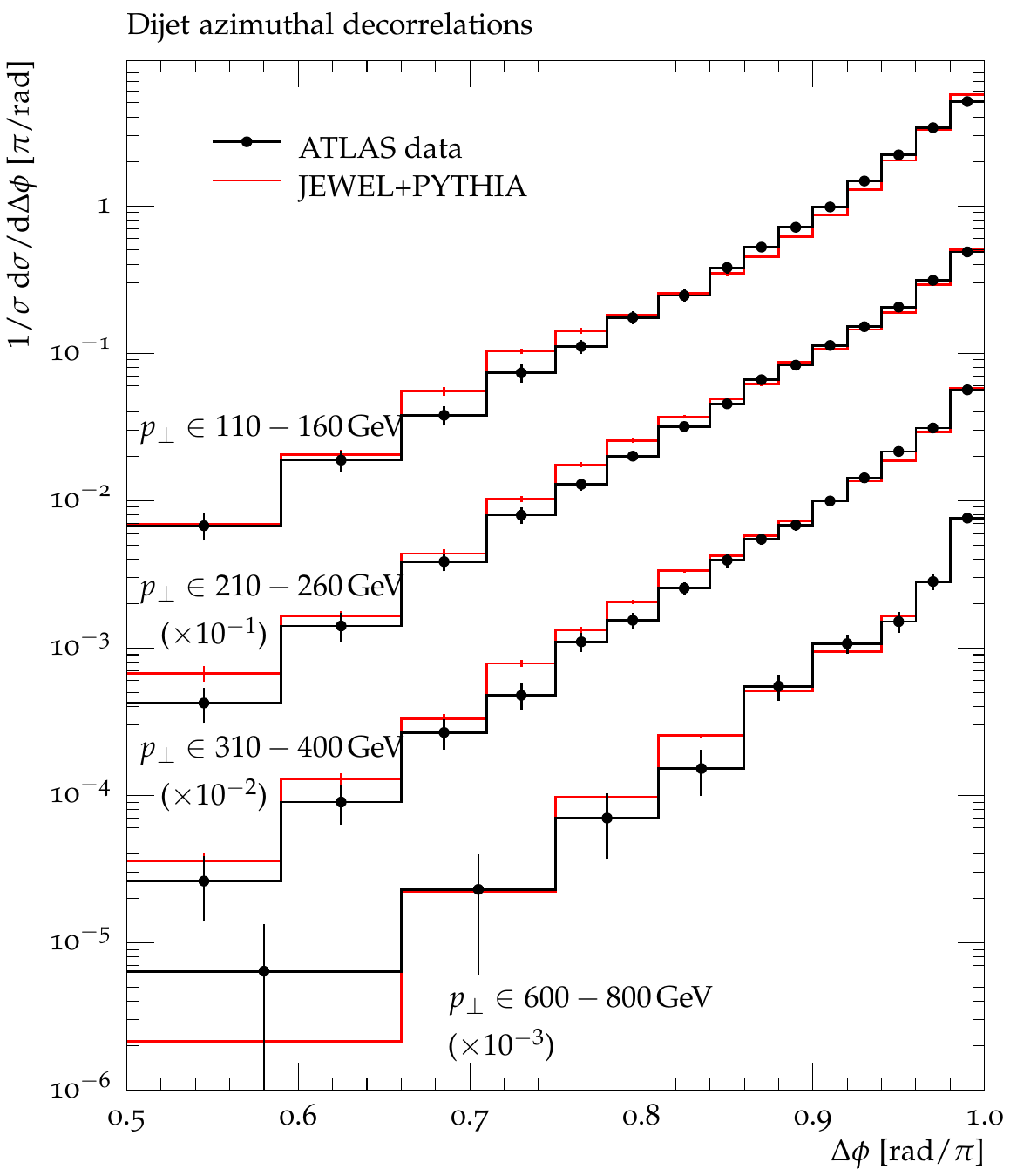}
 \includegraphics[width=0.48\textwidth]{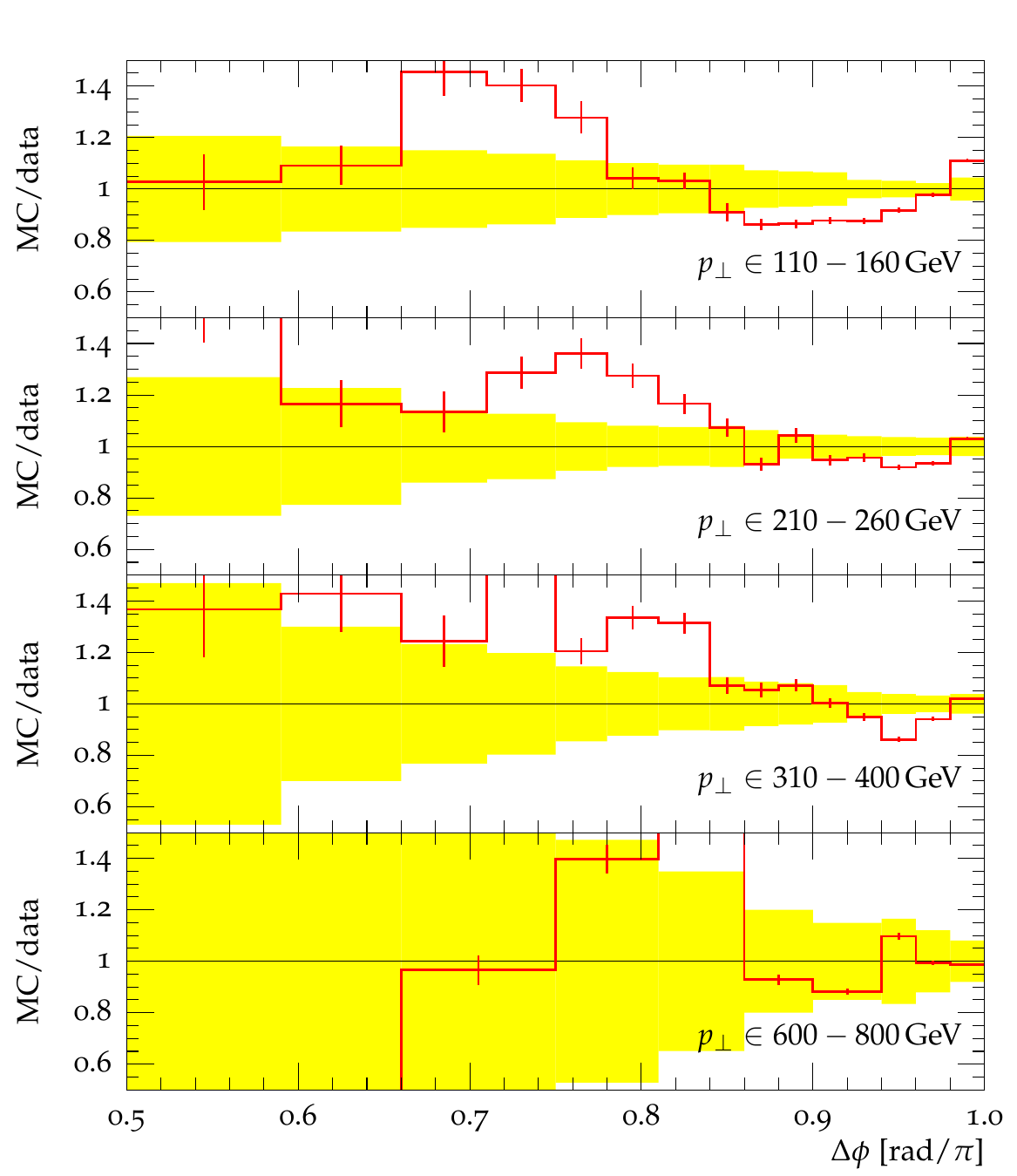}
 \caption{\textsc{Jewel+Pythia} compared to \textsc{Atlas} jet measurements 
   of the azimuthal decorrelation~\cite{daCosta:2011ni} in p+p at 
   $\sqrt{s}=\unit[7]{TeV}$.}
\label{Fig::ppatLHCdecorr}
\end{figure}

At the \textsc{LHC} the comparison of \textsc{Jewel+Pythia} results to the jet 
data is complicated by the underlying event, which is not included in the MC. 
This can, for instance, be seen in the inclusive (track) jet cross section 
shown in \FigRef{Fig::ppatLHCjetspec}: While the agreement between the MC 
results and the data is satisfactory at high $\pt$, the MC falls below the 
data at relatively small transverse momenta, where the contribution from the 
underlying event is largest.  The same effect is visible in the (calorimetric) 
jet shape measurements (\FigRef{Fig::ppatLHCjetshape}), where the 
\textsc{Jewel+Pythia} jet are significantly more collimated at low $\pt$ and 
again the agreement improves with increasing $\pt$.  A complementary 
observable, namely the azimuthal decorrelation shown in 
\FigRef{Fig::ppatLHCdecorr}, is also described reasonably well.

\begin{figure}[t]
 \centering
 \includegraphics[width=0.48\textwidth]{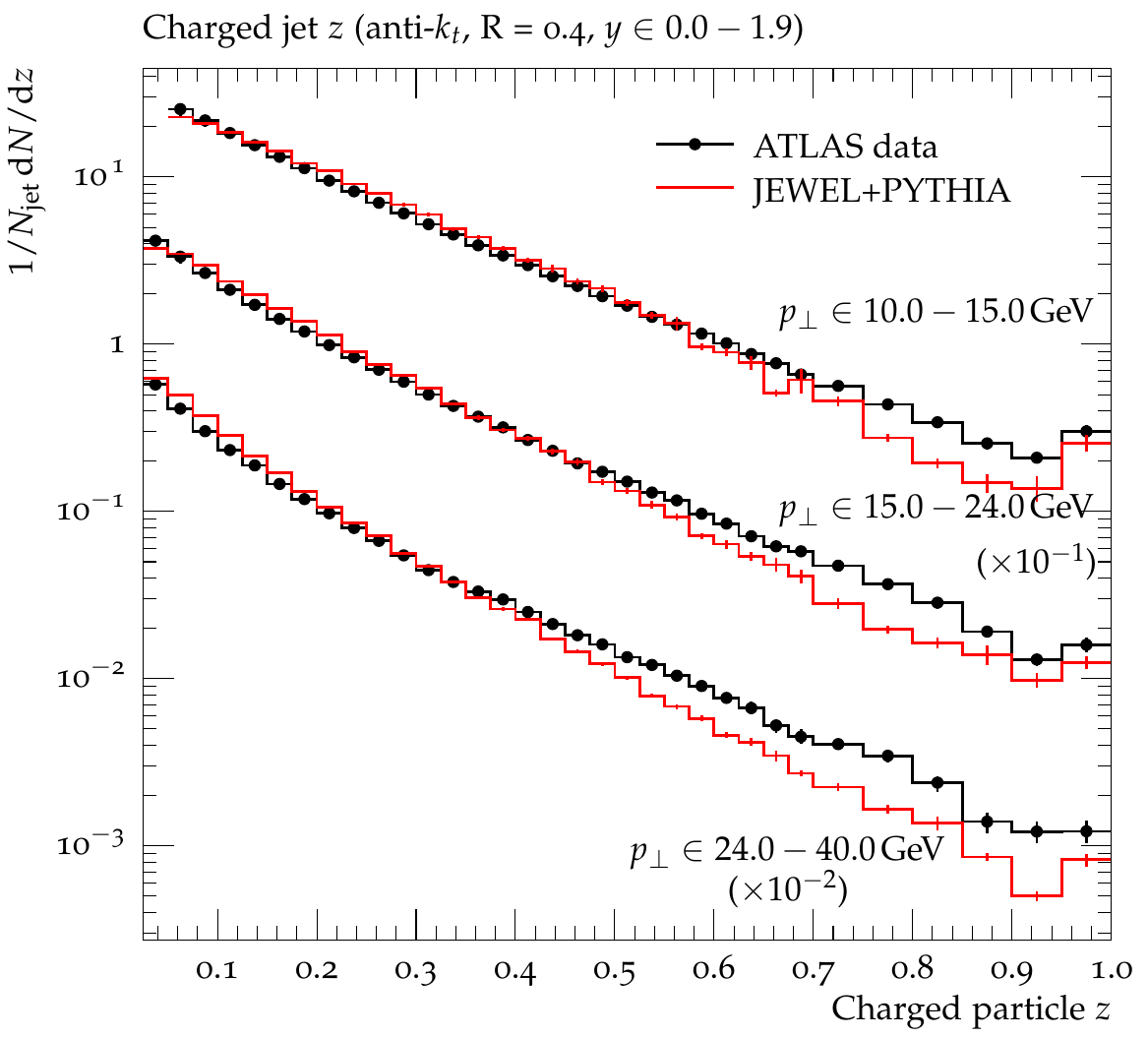}
 \includegraphics[width=0.48\textwidth]{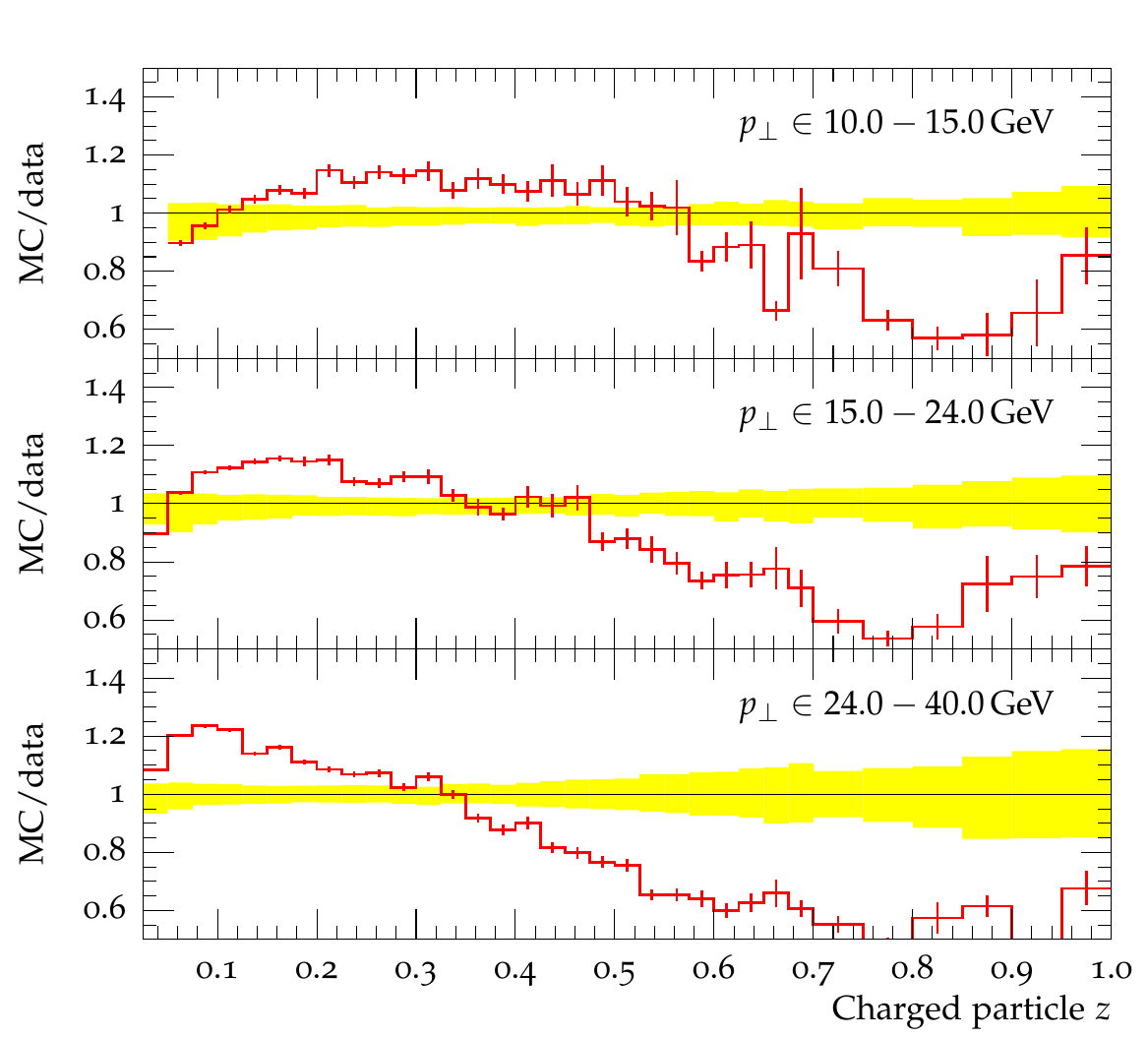}
 \caption{\textsc{Jewel+Pythia} results for the fragmentation function of 
   charged particles in track jets for different jet-$\pt$ bins compared to 
   \textsc{Atlas} data~\cite{Aad:2011gn}.}
\label{Fig::ppatLHCFFz}
\end{figure}

\begin{figure}[t]
 \centering
 \includegraphics[width=0.48\textwidth]{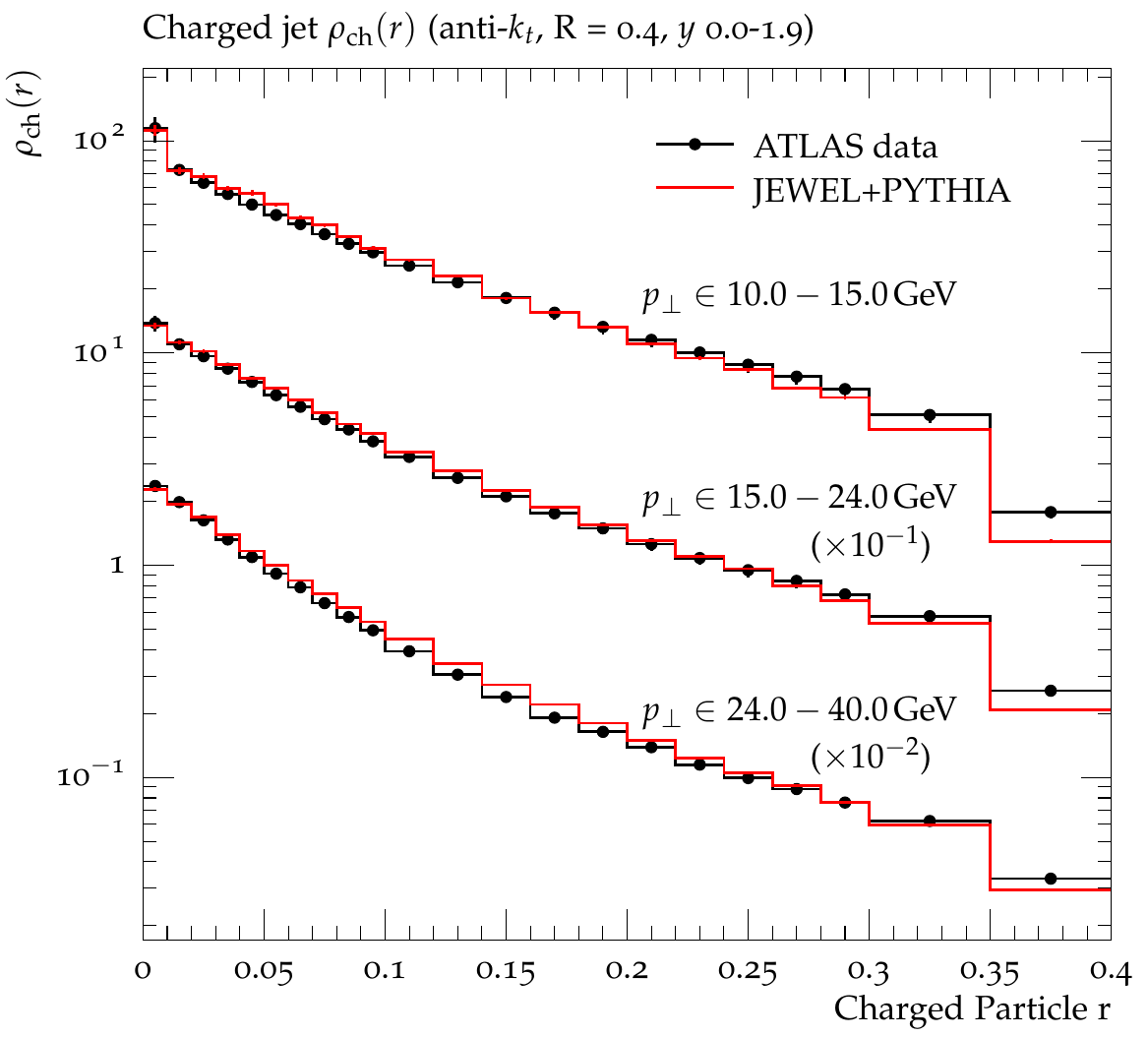}
 \includegraphics[width=0.48\textwidth]{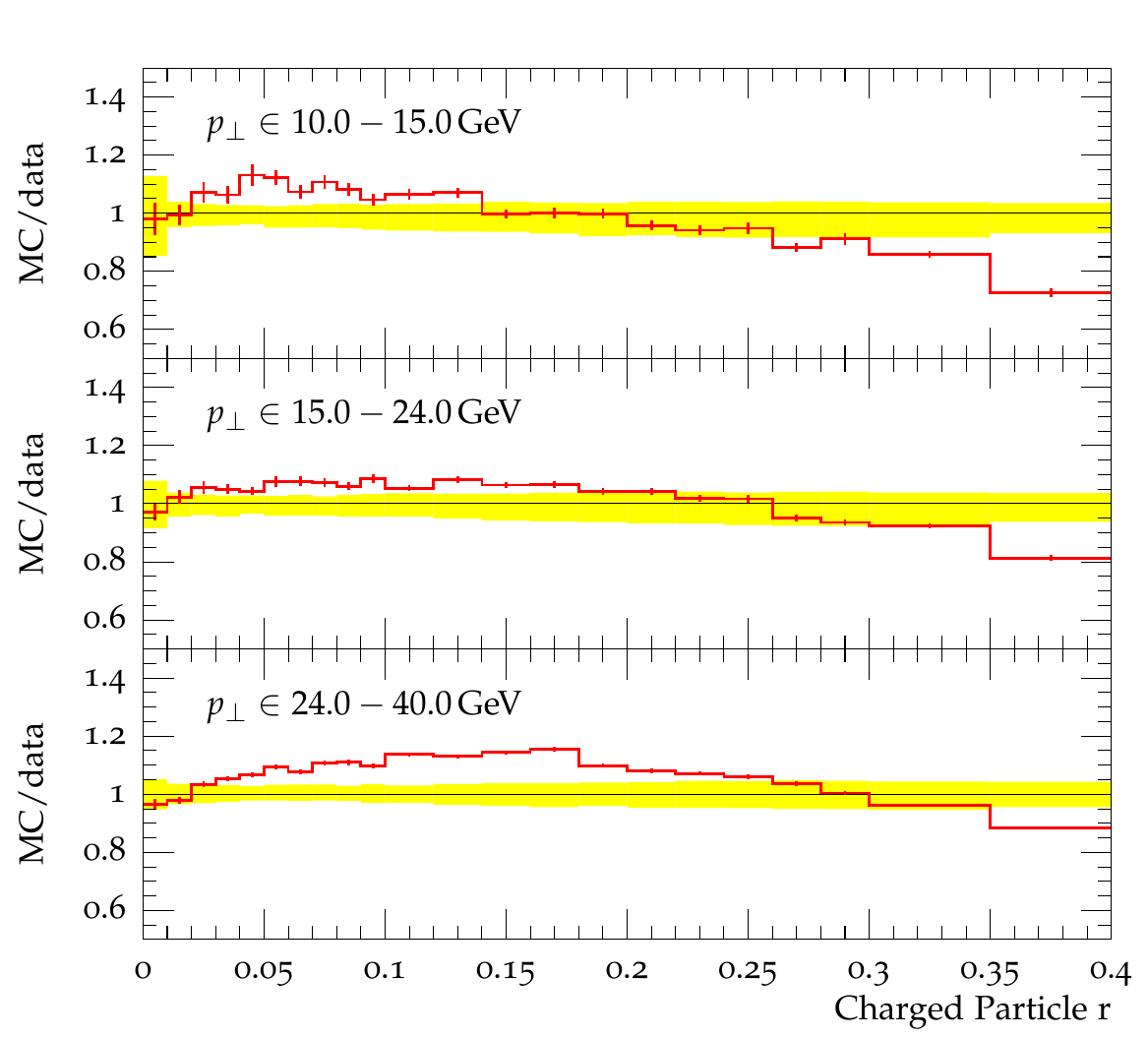}
 \caption{\textsc{Jewel+Pythia} results for the charged particle density 
   $\rho_\text{ch}(r)$ in track jets for different jet-$\pt$ bins compared 
   to \textsc{Atlas} data~\cite{Aad:2011gn}.}
\label{Fig::ppatLHCrho}
\end{figure}

Figure~\ref{Fig::ppatLHCFFz} shows the comparison of \textsc{Jewel+Pythia} to 
the \textsc{Atlas} measurement of the charged particle fragmentation function 
in track jets. Here, the \textsc{Jewel+Pythia} fragmentation consistently is
too soft.  The interpretation of this observation is, however, not 
straightforward, as the measurement is carried out for relatively low 
jet-$\pt$ and it is thus unclear to what extent there is a contamination 
from the underlying event.  Two points hint at an issue in the MC in addition 
to possible underlying event contributions: Firstly, the charged particle 
density inside track jets shown in \FigRef{Fig::ppatLHCrho} shows better 
agreement between MC and data, although also here the components missing at 
large $r$ could be due to underlying event. Secondly, the jet shapes of 
calorimeter jets tend to be slightly too collimated in the MC even at high 
$\pt$ (\FigRef{Fig::ppatLHCjetshape}), which would be consistent with a 
softer fragmentation function.  It should be kept in mind that the jet shape 
$\rho(r)$ is an energy density, while $\rho_\text{ch}(r)$ is a particle density. 

\begin{figure}[t]
 \centering
 \includegraphics[width=0.65\textwidth]{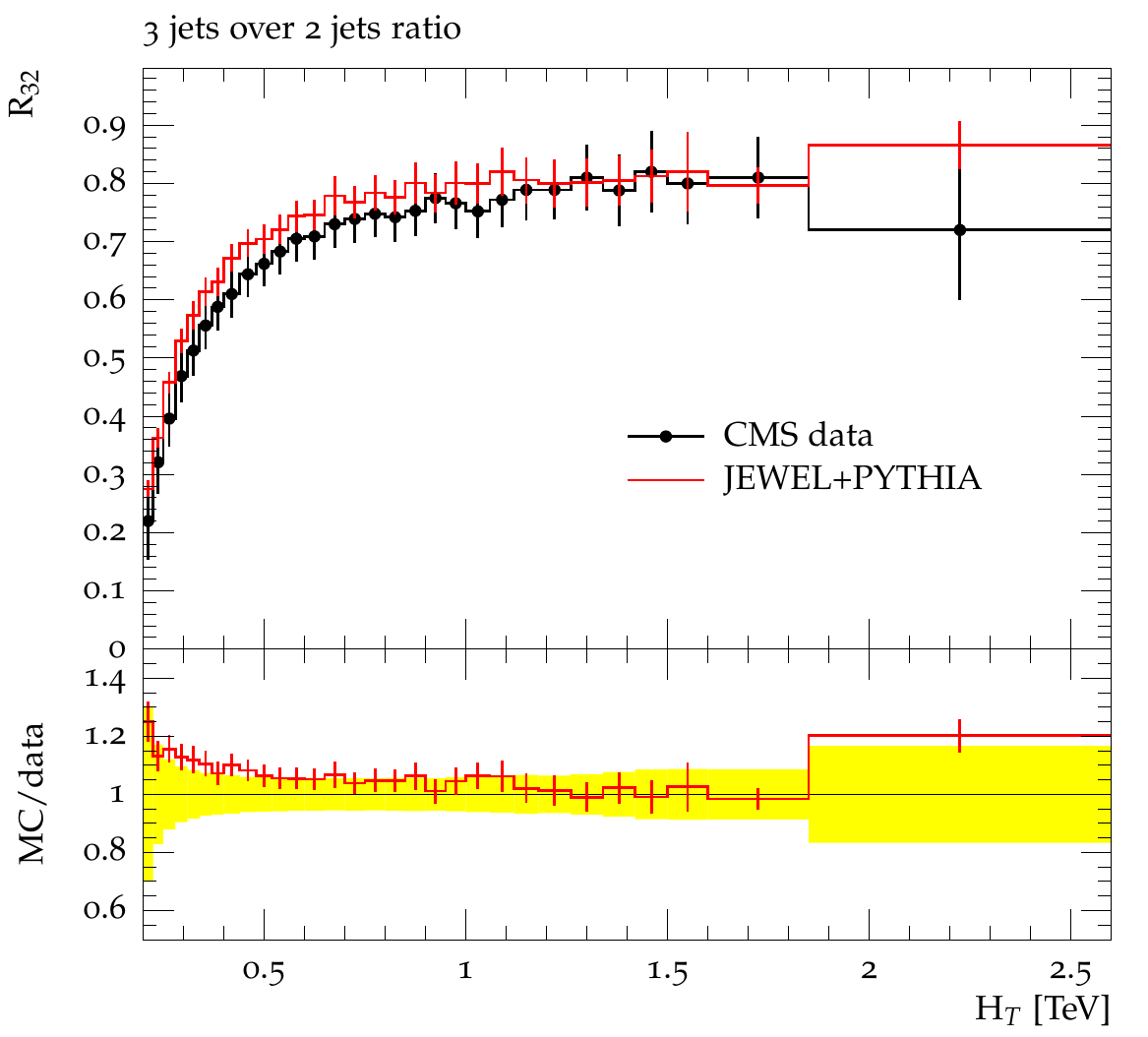}
 \caption{The ratio $R_{32}$ of the inclusive 3-jet to the 2-jet cross section 
   as a function of the scalar sum of the transverse momenta $H_\text{T}$, 
   the anti-$\kt$ jets in this analysis have $\pt > \unit[50]{GeV}$ and 
   $R=0.5$~\cite{Chatrchyan:2011wn}.}
\label{Fig::R32}
\end{figure}

The comparison of the \textsc{Jewel} shower to the measurement of the ratio 
of the 3-jet to the 2-jet cross section by \textsc{CMS} shown in 
Fig.~\ref{Fig::R32} demonstrates that the parton shower provides a very reasonable 
description of the 3-jet matrix elements even well outside the collinear 
region.  This confirms that leading order matrix elements in combination with a 
parton shower are a good approximation to radiative processes and thus provide 
a good estimate of the radiative energy loss.

\begin{figure}[ht]
 \centering
 \includegraphics[width=0.48\textwidth]{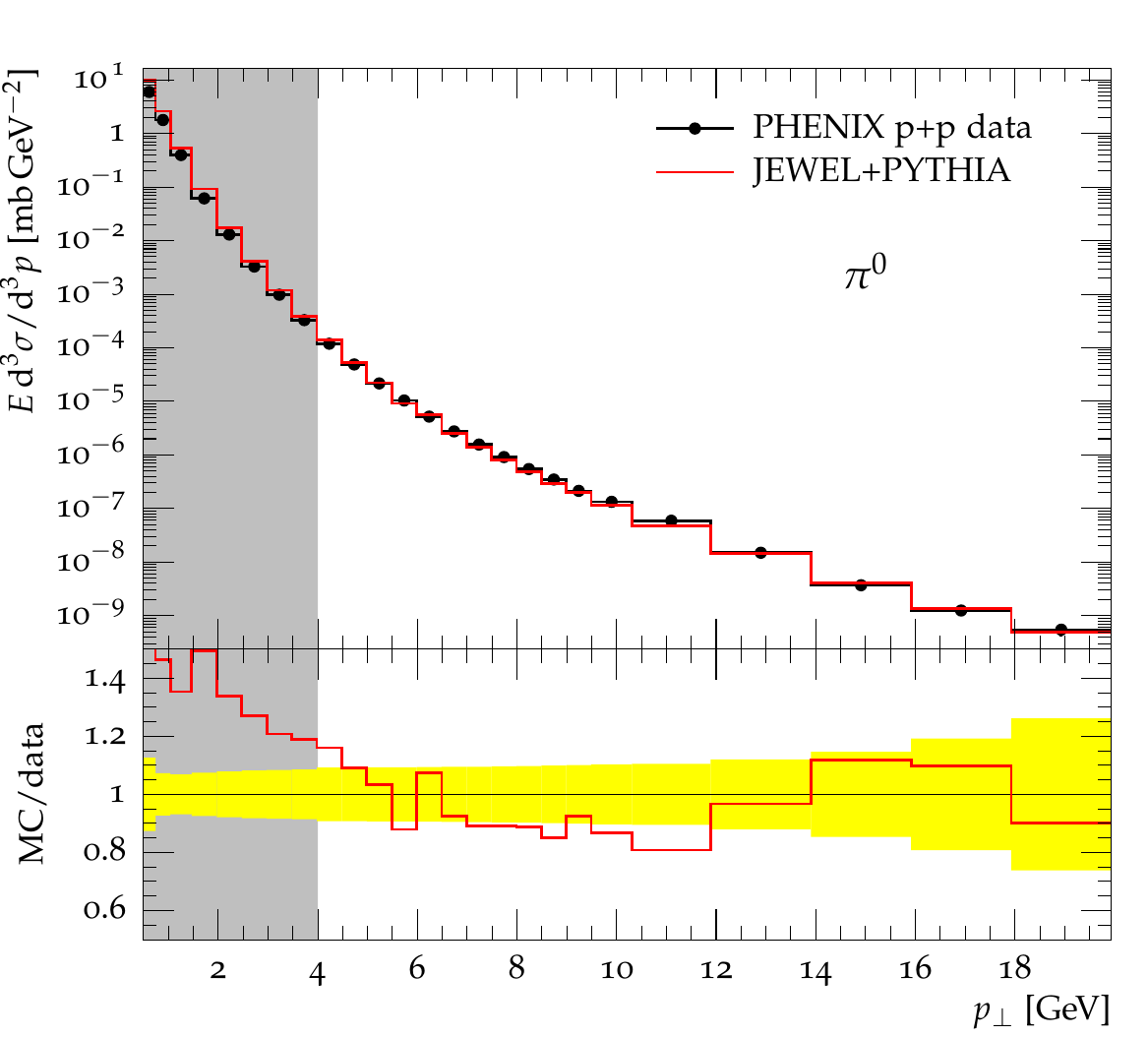}
 \includegraphics[width=0.48\textwidth]{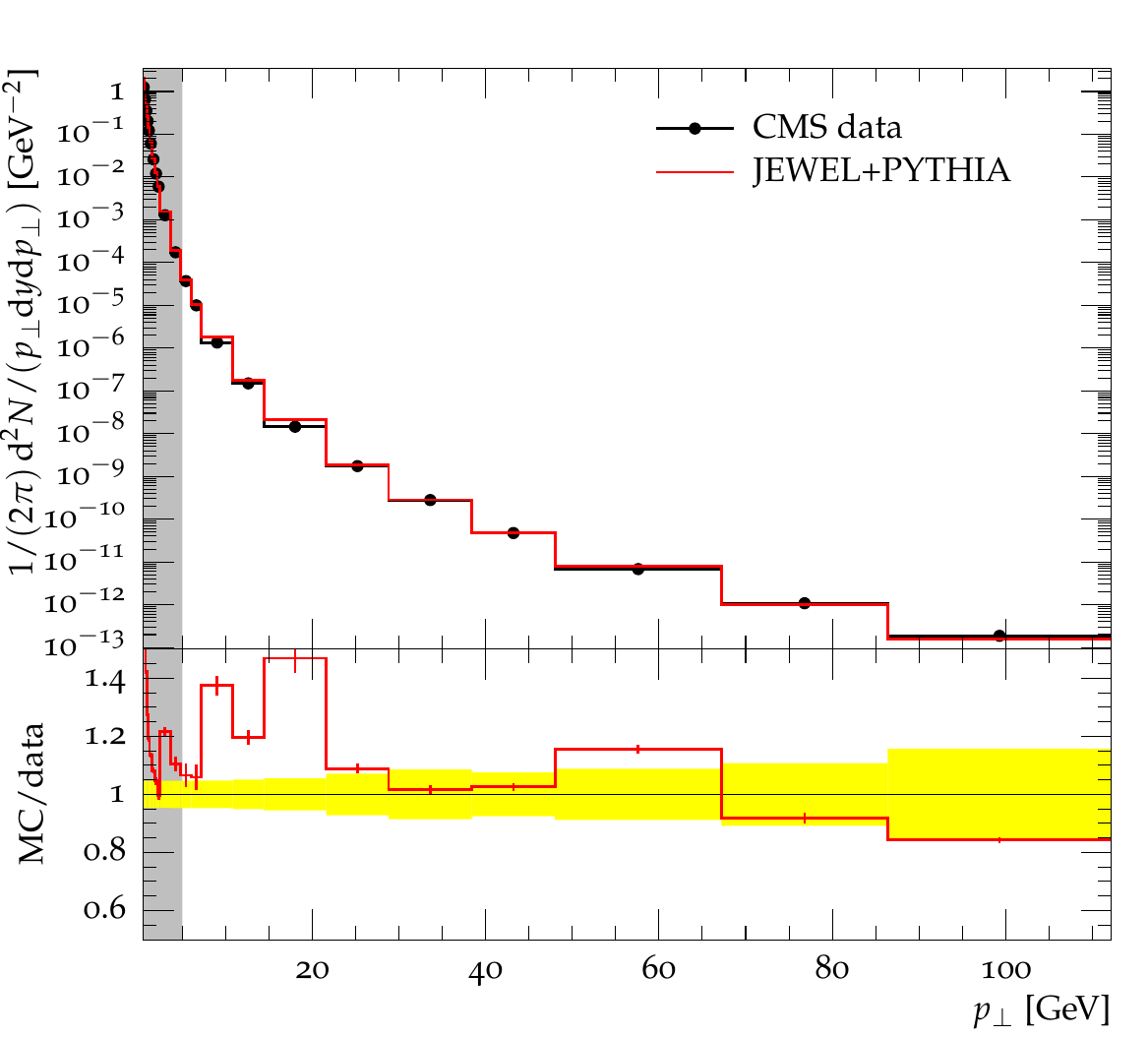}
 \caption{\textbf{LHS:} Neutral pion spectrum in p+p collisions at 
   $\sqrt{s}=\unit[200]{GeV}$ simulated with \textsc{Jewel+Pythia}
   and compared to \textsc{Phenix} data~\cite{Adare:2007dg}.
   \textbf{RHS:} Charged hadron spectrum in p+p collisions at 
   $\sqrt{s}=\unit[2.76]{TeV}$ simulated with \textsc{Jewel+Pythia}
   and compared to \textsc{Cms} data~\cite{CMS:2012aa}.}
 \label{fig::pi0spec}
\end{figure}

Fig.~\ref{fig::pi0spec} shows a comparison with the neutral pion spectrum 
measured by \textsc{Phenix} and the charged particle spectrum measured by 
\textsc{CMS}, which form the baseline for the respective measurements of the 
nuclear modification factor $R_\text{AA}$.  In the former case, above 
$\pt \simeq \unit[4]{GeV}$, the \textsc{Jewel+Pythia} results agree with 
the data on a level of roughly \unit[15]{\%} over about 6 decades and 
thus provide a realistic baseline for the $R_\text{AA}$ determination. 
At \textsc{LHC} energies the hadron spectrum seems to be slightly softer in 
the simulation than in data.  However, deviations of this size have only a 
very small effect on the nuclear modification factor. The uncertainty arising 
from this is much smaller than, for instance, the uncertainty arising from 
the choice of infrared regulator $\mu_\textbf{D}$.

\FloatBarrier

\subsection{Hadron suppression}
\begin{figure}[t]
 \centering
 \includegraphics[angle=-90,width=0.65\textwidth]{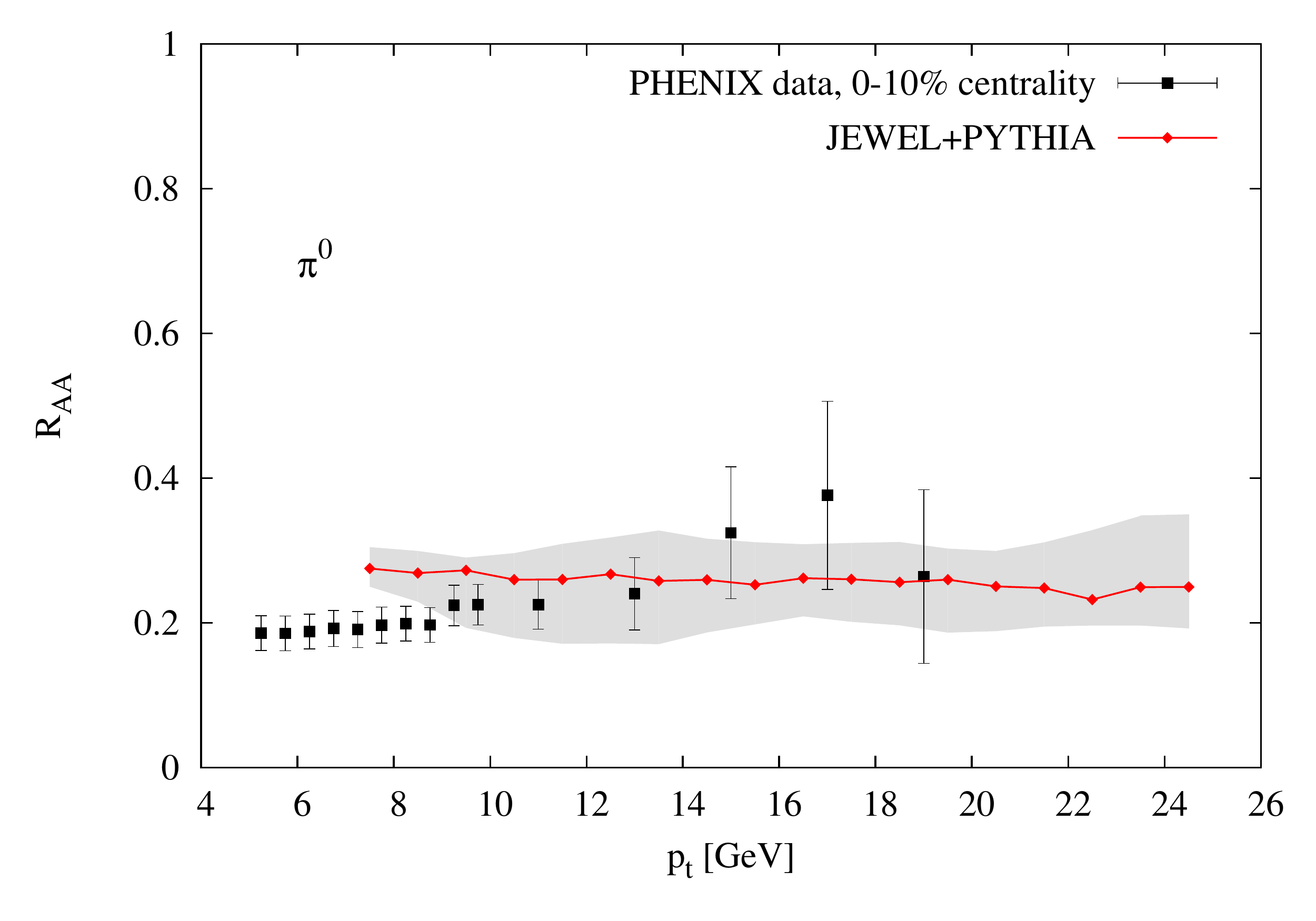}
 \caption{Nuclear modification factor for neutral pions in Au+Au collisions at 
   $\sqrt{s}=\unit[200]{A\,GeV}$ in the \unit[0-10]{\%} centrality class 
   simulated with \textsc{Jewel+Pythia} and compared to \textsc{Phenix}
   data~\cite{Adare:2012wg} (only systematic errors shown, statistical 
   errors are smaller than the systematic ones everywhere). The grey band 
   indicates a variation of the Debye mass by $\pm\unit[10]{\%}$.}
 \label{fig::phenixraa}
\end{figure}

To obtain a fair agreement with the measured nuclear modification factor at 
\textsc{RHIC}, \textsc{Jewel+Pythia} requires an initial temperature of 
$\ti=\unit[350]{MeV}$ at initial proper time $\taui=\unit[0.8]{fm}$, see 
Fig.~\ref{fig::phenixraa}. This is remarkably consistent with the input 
parameters in fluid dynamic simulations of heavy ion collisions~\cite{Soltz:2012rk,Csanad:2011jq,Nayak:2008tk}. 
We note that at high $\pt$, where the Monte Carlo results are reliable, they 
reproduce both the factor $\sim 5$ suppression and the approximately flat 
$\pt$-dependence seen in data.  Varying the Debye mass by $\pm \unit[10]{\%}$ 
has a significant effect on the overall suppression, indicated by the grey 
band, but hardly affects the shape. The dependence on the exact choice of 
the regulator is clearly sizeable.  The pragmatic approach is to regard it 
as a parameter, fix it together with the medium parameters at one point 
(in this case the neutral pion suppression at \textsc{RHIC}) and use the 
same value for all calculations.  But while this may be a practical working 
solution it does not address the underlying problem.

\begin{figure}[t]
 \centering
 \includegraphics[angle=-90,width=0.65\textwidth]{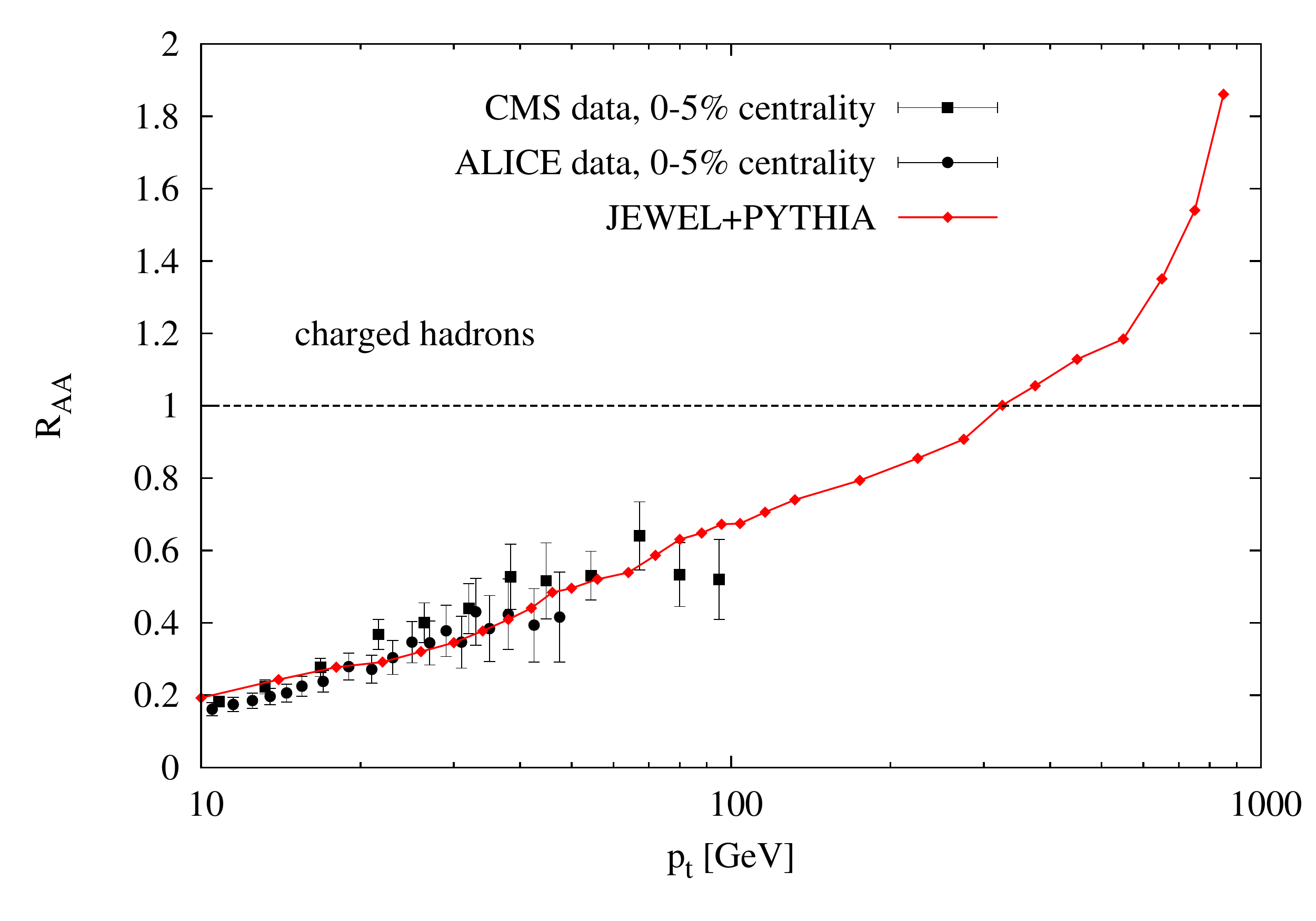}
 \caption{Nuclear modification factor for charged hadrons in Pb+Pb collisions 
   at $\sqrt{s}=\unit[2.76]{A\,TeV}$ in the \unit[0-5]{\%} centrality class 
   simulated with \textsc{Jewel+Pythia} and compared to preliminary 
   \textsc{CMS}~\cite{CMS:2012aa} and 
   \textsc{Alice}~\cite{:2012eq} data.}
 \label{fig::cmsraa}
\end{figure}

The charged hadron multiplicities measured in heavy ion collisions constrain 
the initial entropy density of the system, $s_{\text{i}} \taui \propto \d N/\d 
y$, $s_{\text{i}} \propto \epsilon_{\text{i}}/\ti \propto \ti^3$ and therefore 
allow to relate the initial temperatures at \textsc{RHIC} and at the 
\textsc{LHC},
\begin{equation}
 \ti^{\text{LHC}} = \ti^{\text{RHIC}} \left( 
\frac{\taui^{\text{RHIC}}}{\taui^{\text{LHC}}} \frac{\left.
  {\d N / \d y} \right|_{\text{LHC}}}{\left. {\d N /\d 
y}\right|_{\text{RHIC}}} \right)^{1/3} \,. 
\end{equation}
The observation of a factor 2.2 increase in the event multiplicity from 
\textsc{RHIC} to \textsc{LHC} is therefore consistent with an initial 
temperature $\ti=\unit[530]{MeV}$ at $\taui=\unit[0.5]{fm}$ at the 
\textsc{LHC}. There is some freedom in initializing the fluid dynamic 
evolution at the \textsc{LHC} at a different initial time $\taui$, but this is 
numerically unimportant.  At early times the parton shower is dominated by 
emissions at rather high scales initiated by the initial hard scattering, and 
this high virtuality protects the partons from medium--\-induced emissions and 
makes them insensitive to the medium at early times. Thus, the medium at the 
\textsc{LHC} is specified in terms of parameters fixed in 
Fig.~\ref{fig::phenixraa}.  As seen from Fig.~\ref{fig::cmsraa}, the 
calculation of \textsc{Jewel+Pythia} then leads to a very good agreement 
with preliminary data of the nuclear modification factor at the \textsc{LHC}
without any additional adjustments.  This is a remarkable success of our model.

To understand the characteristically different $\pt$-dependencies of  
$R_{\text{AA}}$ at \textsc{RHIC} and at the \textsc{LHC}, we have performed 
control simulations in which a \textsc{LHC}-like distribution of hard 
processes is fragmented in a \textsc{RHIC}-like medium, and vice versa (data 
not shown).  This showed that the significant change in the slope of  
$R_{\text{AA}}(\pt)$ from \textsc{RHIC} to the \textsc{LHC} can be attributed 
fully to the $\sqrt{s}$-dependence of the distribution of initial hard 
processes.  We therefore conclude that a purely perturbative dynamics of parton 
energy loss supplemented by an arguably simple model of the medium whose 
characterisation matches physical expectations, can account for the main 
features of the measured nuclear modification factors, including the strength 
of the suppression pattern, and its $\sqrt{s}$- and $\pt$-dependence. 

At very large $\pt$ the nuclear modification factor continues to rise above 
unity. This is a purely kinematical effect that becomes visible at very large 
$\pt$ where the energy loss starts to vanish. The elastic scattering of 
energetic partons converts longitudinal into transverse momentum turning the  
$\pt$-spectrum harder. While this is a generic effect its size and turn-on  
point will to some degree depend on the medium model, as they are sensitive 
to the amount of scattering centres encountered in the forward direction. 
This effect would not be accessible in the standard eikonal treatment of 
parton energy loss that neglects the change in propagation direction of the 
projectile parton.  We believe it is thus a generic testable prediction of 
our model.

\FloatBarrier

\subsection{Modification of jets}

\begin{figure}[t]
 \centering
 \includegraphics[width=0.48\textwidth]{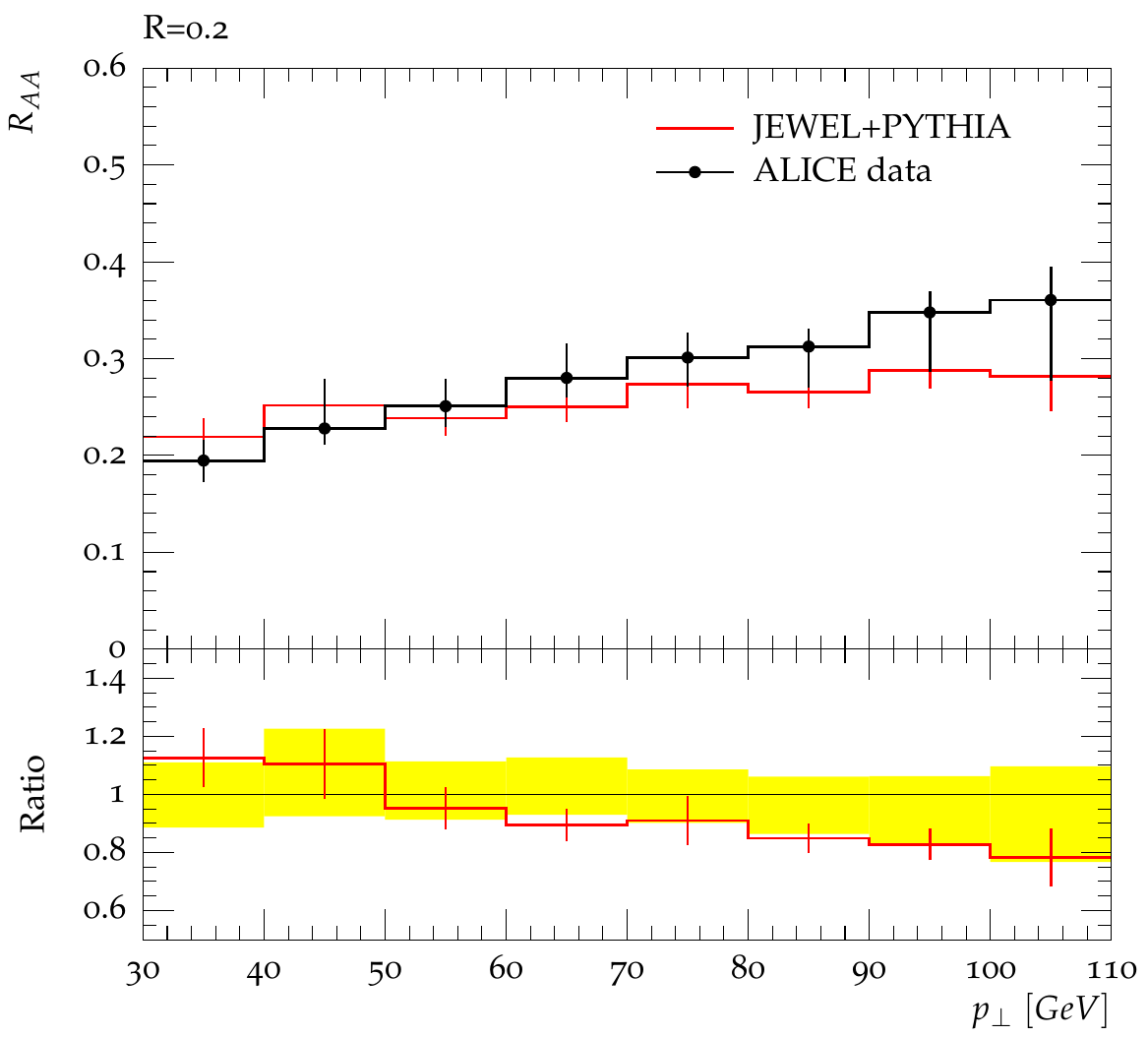}
 \includegraphics[width=0.48\textwidth]{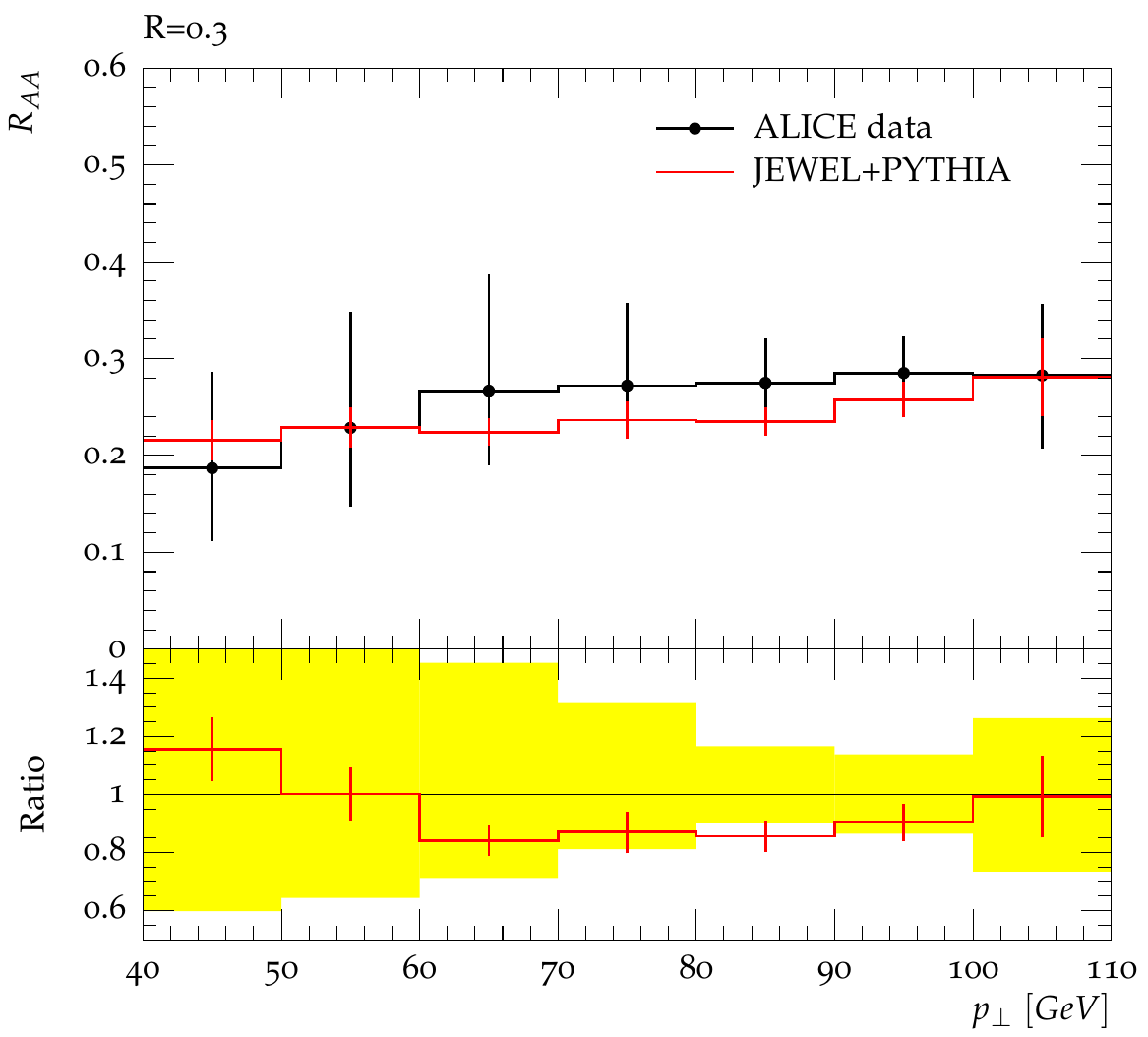}
 \caption{\textsc{Jewel+Pythia} results for $R_\text{AA}$ of jets in 
  Pb+Pb collisions at $\sqrt{s_\text{NN}} = \unit[2.76]{TeV}$ compared to 
  \textsc{Alice} data~\cite{:2012ch} for two values of the jet radius 
  (correlated systematic errors not shown).}
\label{Fig::JetRAA}
\end{figure}

\begin{figure}[t]
 \centering
 \includegraphics[width=0.48\textwidth]{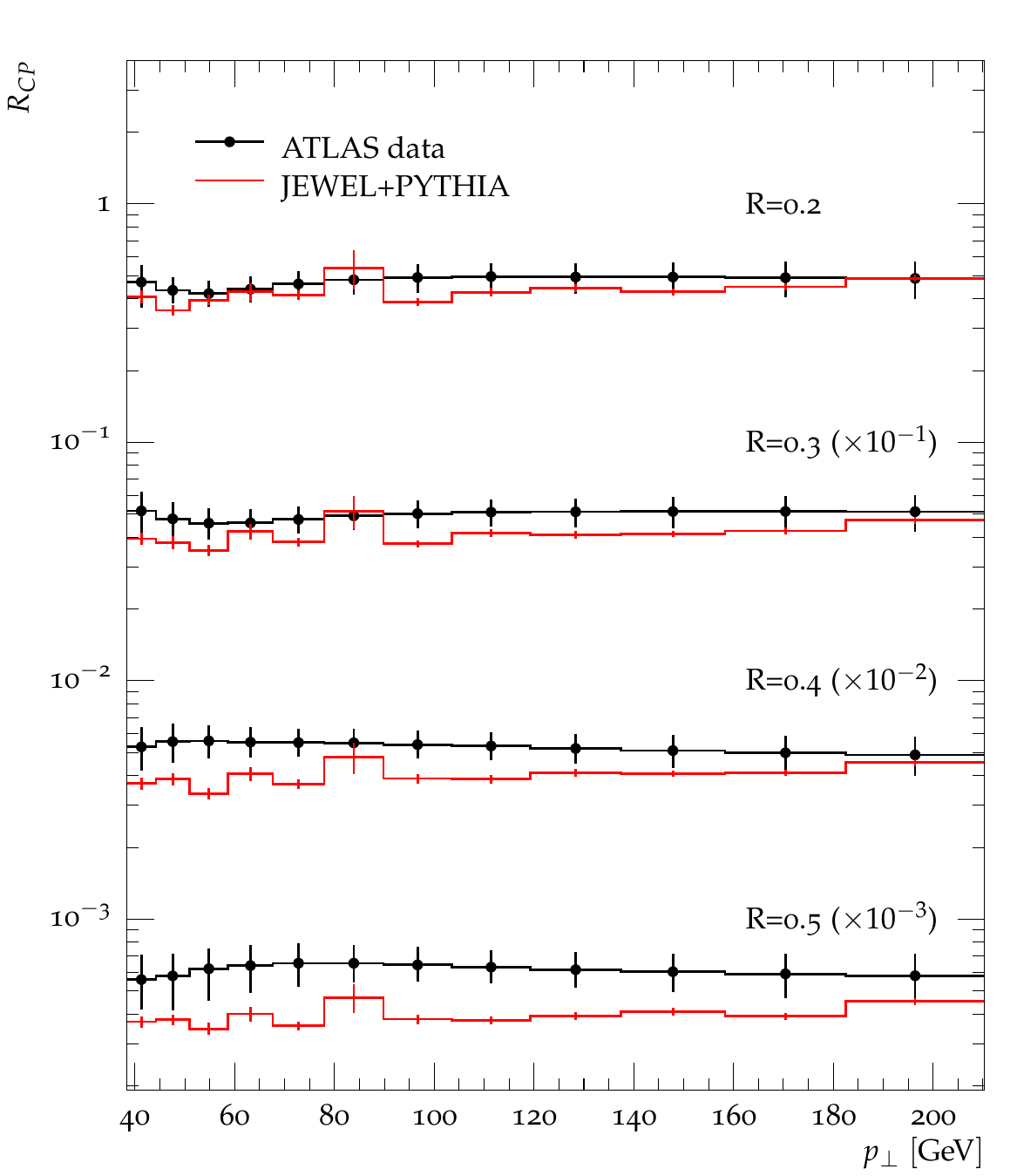}
 \includegraphics[width=0.48\textwidth]{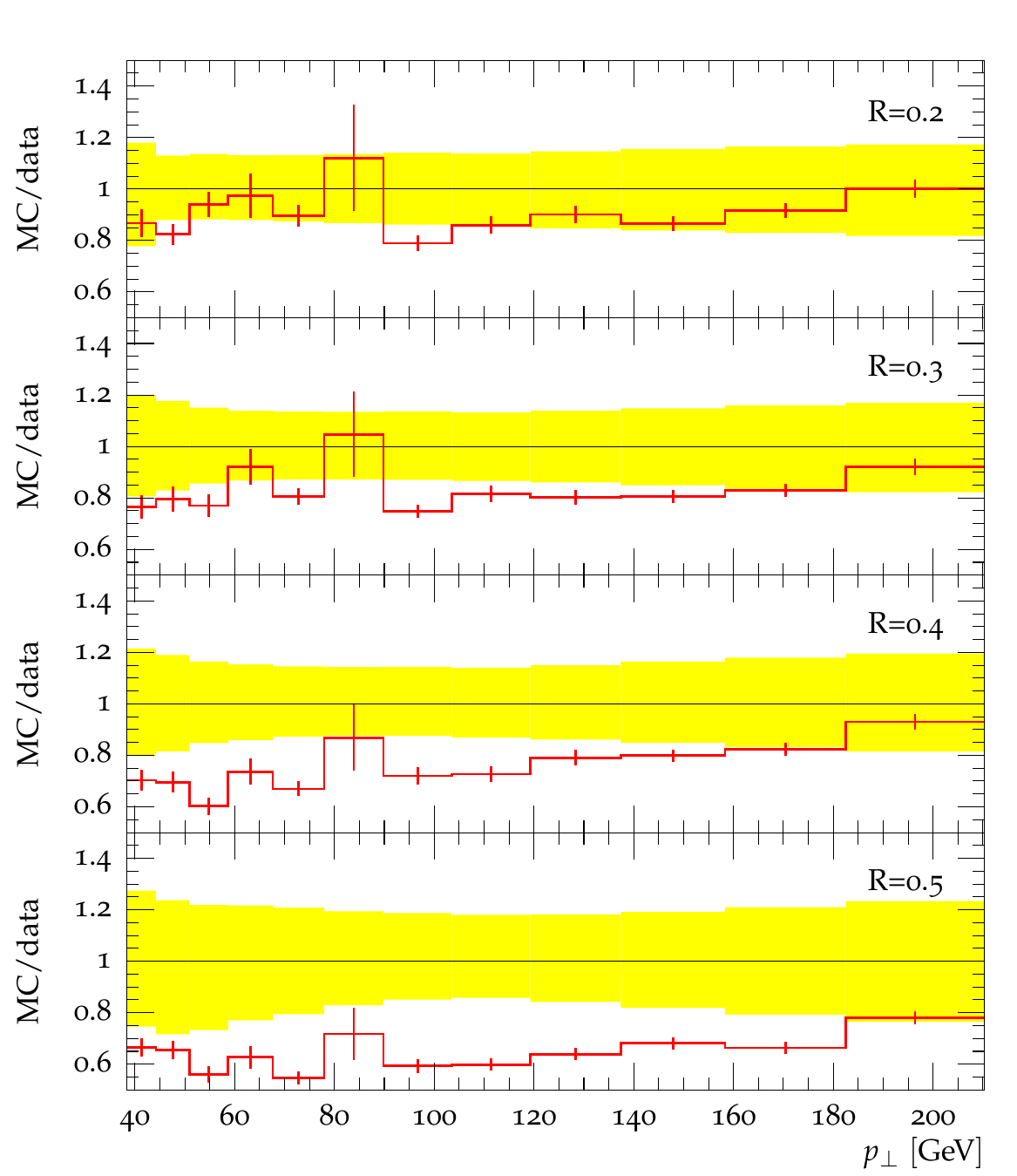}
 \caption{\textsc{Jewel+Pythia} results for $R_\text{CP}$ of jets in 
   Pb+Pb collisions at $\sqrt{s_\text{NN}} = \unit[2.76]{TeV}$ compared to 
   \textsc{Atlas} data~\cite{:2012is} for different values of the jet radius. 
   The ratio is taken between the \unit[0-10]{\%} and the \unit[60-80]{\%} 
   centrality class.}
\label{Fig::JetRCP}
\end{figure}

Analogously to the single-inclusive hadron suppression discussed in the 
previous section also the single-inclusive jet suppression has been measured 
by \textsc{Alice} and \textsc{Atlas}. In \FigRef{Fig::JetRAA} and 
\ref{Fig::JetRCP} the experimental results for different values of the jet 
radius are compared to \textsc{Jewel+Pythia}. The \textsc{Alice} results 
(which are relative to a pp baseline) and the \textsc{Atlas} results 
(which are relative to peripheral Pb+Pb spectra) for small jet radii are very 
well reproduced, but there is some discrepancy in the \textsc{Atlas} results 
for large jet radii.  Also, in the $R=0.2$ measurement by \textsc{Alice}, the 
$\pt$-dependence seems to be weaker in the MC than in the data. 
There is no indication of a significant rise of the jet suppression at large 
$\pt$ in the simulation, which shows that jets can behave differently from leading hadrons. 
The fact that 
the disagreement between \textsc{Jewel+Pythia} and the \textsc{Atlas} data 
increases with the jet radius $R$ could hint at an issue with the background 
subtraction. 

For the simulation of the single-inclusive hadron spectra the recoiling 
scattering centres are kept in the event and hadronised together with the 
jets.  In the case of jet analyses, however, the recoils have to be taken out, 
since the experimental jet reconstruction involves the subtraction of 
background.  These two procedures are not exactly identical for several 
reasons.  Firstly, in \textsc{Jewel+Pythia} the hadronic final state is not 
the incoherent sum of a jet component and a contribution from the recoils, 
due to the colour connections between the jet partons and the recoiling 
scattering centres.  Secondly, in the experimental jet reconstruction 
procedure, only a background estimate based on the average activity in the 
event can be subtracted.  While to some degree it is possible to deal with 
background fluctuations using unfolding techniques~\cite{Cacciari:2010te,
Cacciari:2011tm,deBarros:2012ws}, the correlations between 
the jet and the background cannot be assessed in this way.  There will thus 
always remain a residual uncertainty when comparing any MC results to jet 
data.  Clearly, along the same line of reasoning, the background subtraction 
becomes more involved as the jet radius increases and the observation 
that the discrepancy between \textsc{Jewel+Pythia} and data increases with 
jet radius may hint at a problem in this direction.

\smallskip

\begin{figure}[t]
 \centering
 \includegraphics[width=0.48\textwidth]{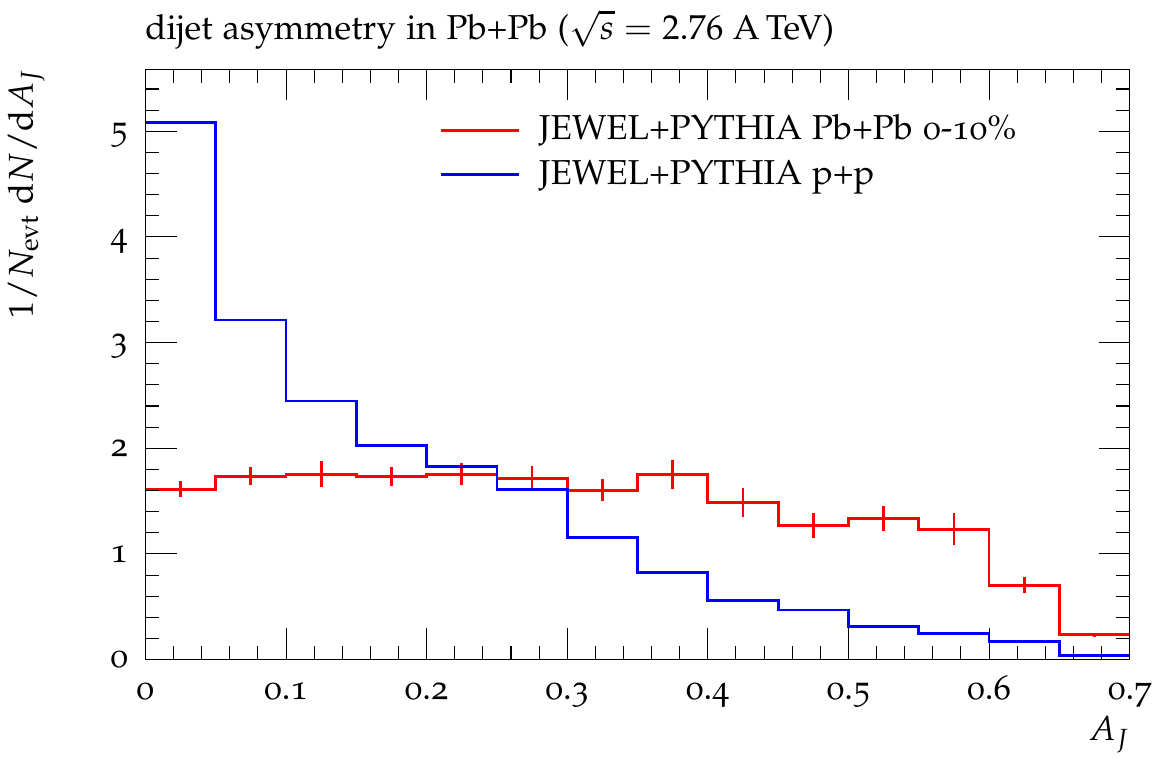}
 \includegraphics[width=0.48\textwidth]{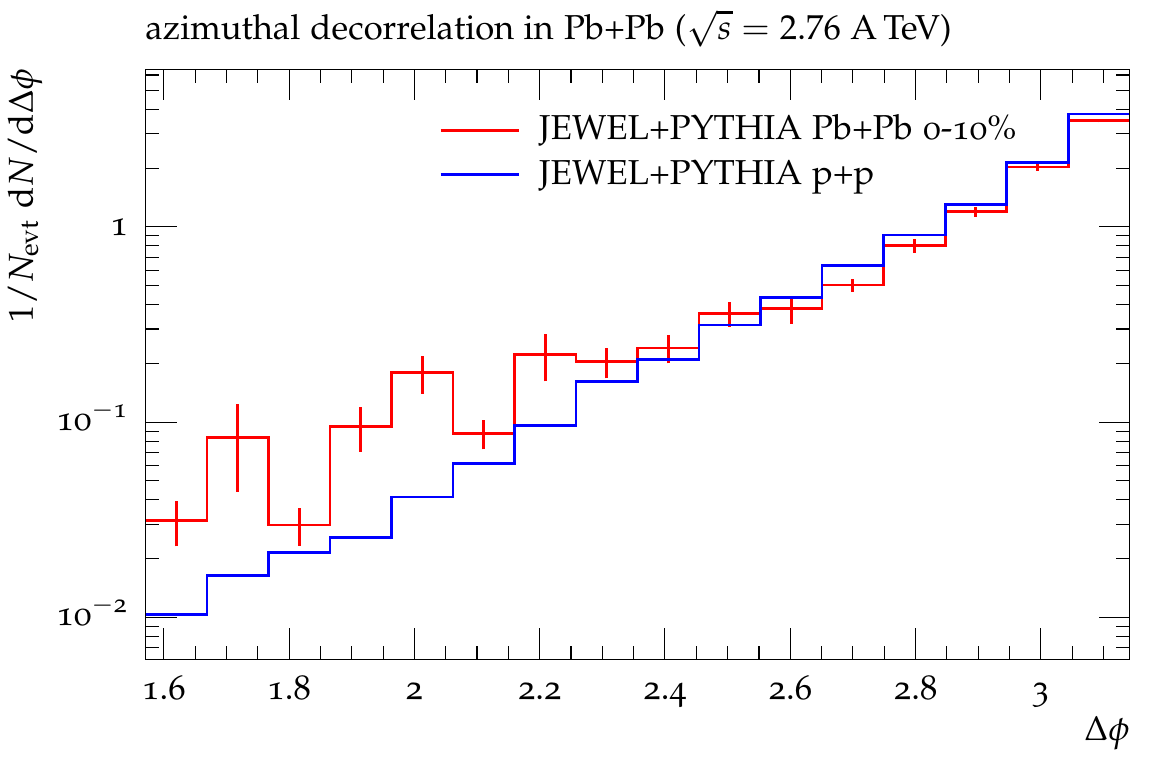}
 \caption{Di-jet asymmetry $A_J = (E_{\perp,1}-E_{\perp,2})/(E_{\perp,1}+E_{\perp,2})$ 
   and azimuthal decorrelation in central Pb+Pb and p+p collisions at 
   $\sqrt{s_\text{NN}} = \unit[2.76]{TeV}$, the cuts are the same as in the 
   \textsc{Atlas} analysis~\cite{Aad:2010bu}.
}
\label{Fig::AJ}
\end{figure}

Both \textsc{Atlas}~\cite{Aad:2010bu} and \textsc{CMS}~\cite{Chatrchyan:2012gw}
have measured the di-jet $\pt$-asymmetry $A_J$ and the azimuthal decorrelation 
in central Pb+Pb events. The \textsc{Jewel+Pythia} results obtained with the 
\textsc{Atlas} cuts are shown in Fig.~\ref{Fig::AJ}.  A direct quantitative
comparison to either \textsc{Atlas} or \textsc{Cms} data is not possible, as 
the data are not unfolded for the jet energy resolution, which is known to have
a sizeable effect especially on the shape of the di-jet asymmetry.  Qualitatively, 
however, the behaviour of the \textsc{Jewel+Pythia} results resembles 
observations made in the data: There is a significant broadening of the $A_J$ 
distribution in central Pb+Pb events as compared to the p+p baseline. The 
azimuthal decorrelation, on the other hand, hardly changes; there is only a 
mild increase at separations $\Delta \phi \sim \pi/2$.

\smallskip

\begin{figure}[t]
 \centering
 \includegraphics[width=0.48\textwidth]{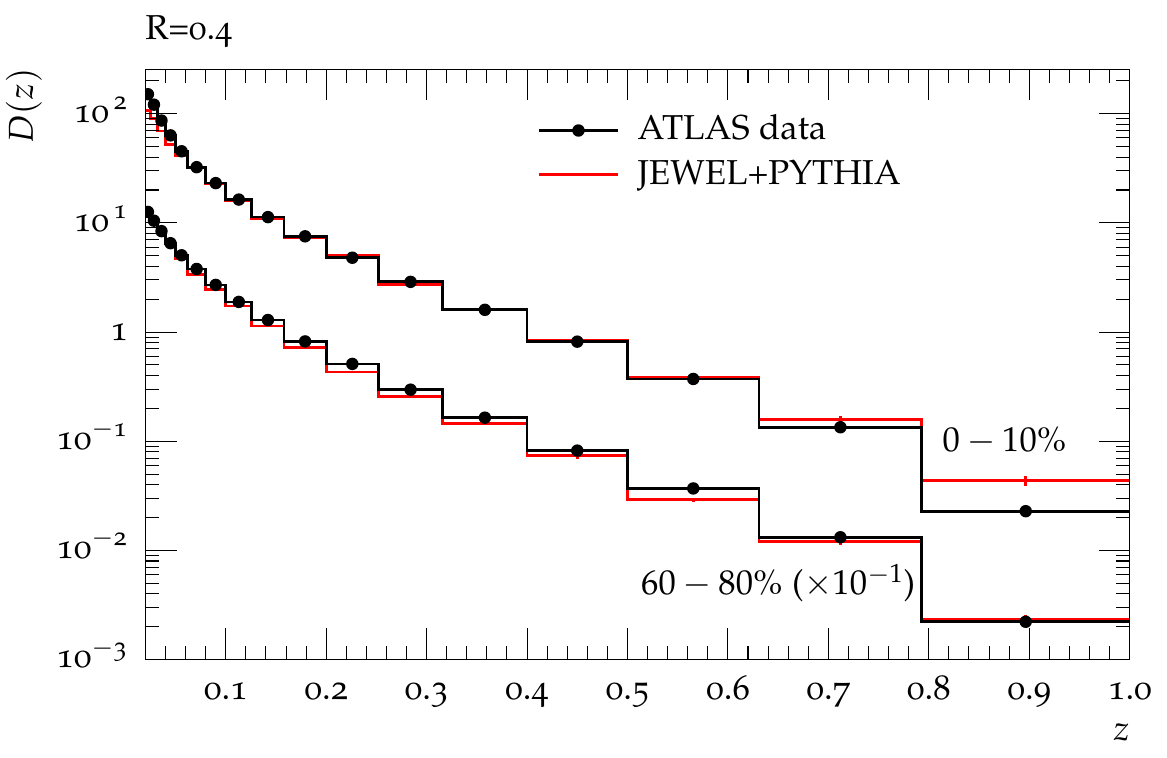}
 \includegraphics[width=0.48\textwidth]{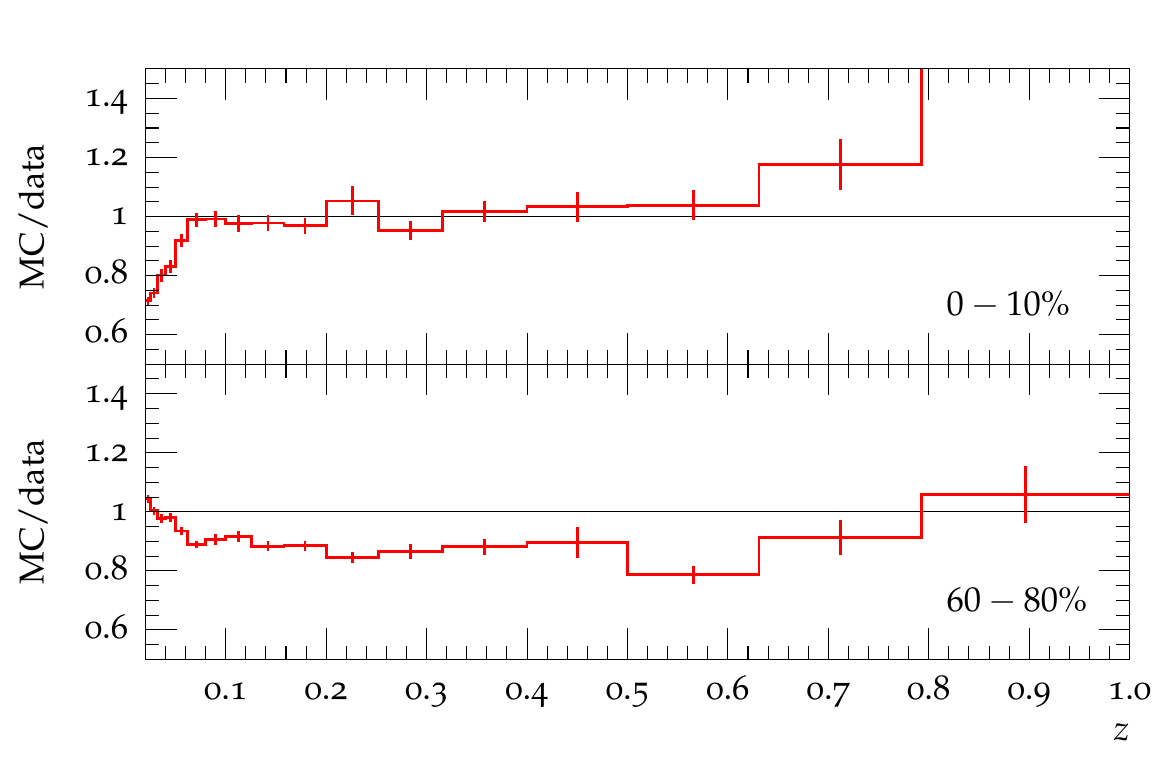}
 \caption{\textsc{Jewel+Pythia} results for the fragmentation function 
   $D(z)$ in peripheral and central Pb+Pb events compared to 
   \textsc{Atlas} data~\cite{ATLAS-CONF-2012-115} 
   (data points read off the plots, no errors shown).}
\label{Fig::FFz}
\end{figure}

\begin{figure}[t]
 \centering
 \includegraphics[width=0.48\textwidth]{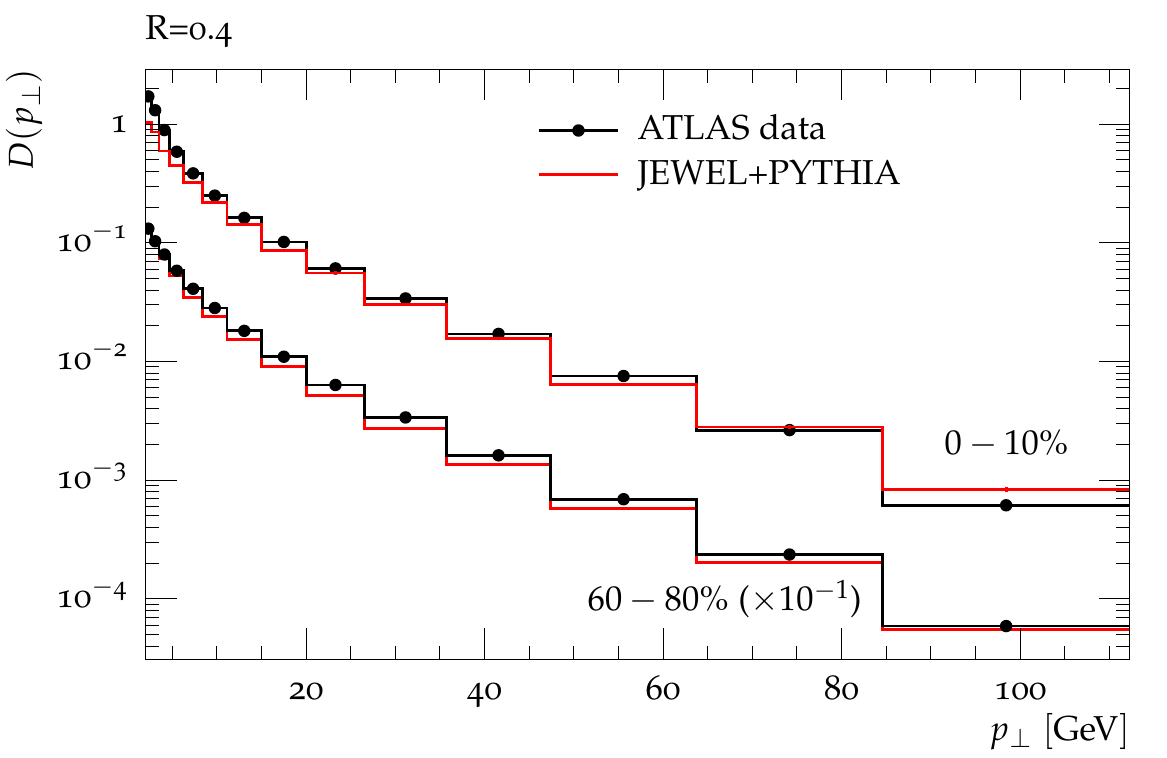}
 \includegraphics[width=0.48\textwidth]{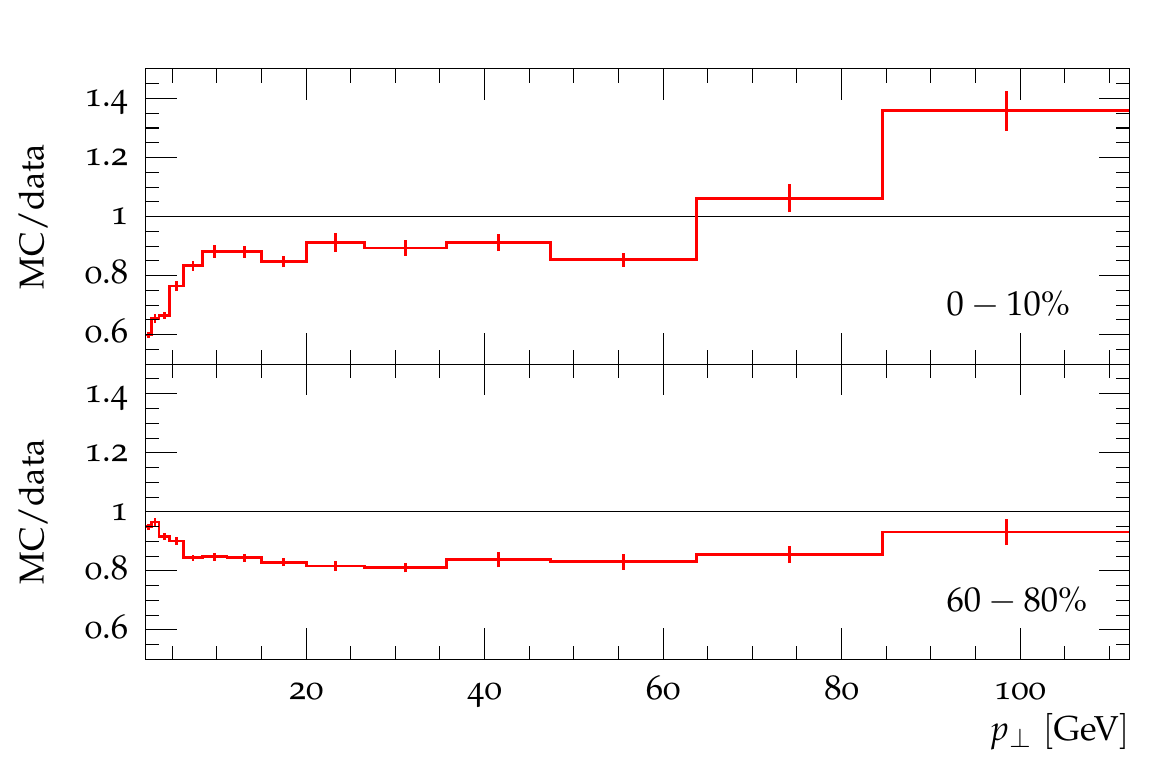}
 \caption{\textsc{Jewel+Pythia} results for the fragmentation function 
   $D(\pt)$ in peripheral and central Pb+Pb events compared to 
   \textsc{Atlas} data~\cite{ATLAS-CONF-2012-115} 
   (data points read off the plots, no errors shown).}
\label{Fig::FFpt}
\end{figure}

\begin{figure}[t]
 \centering
 \includegraphics[width=0.48\textwidth]{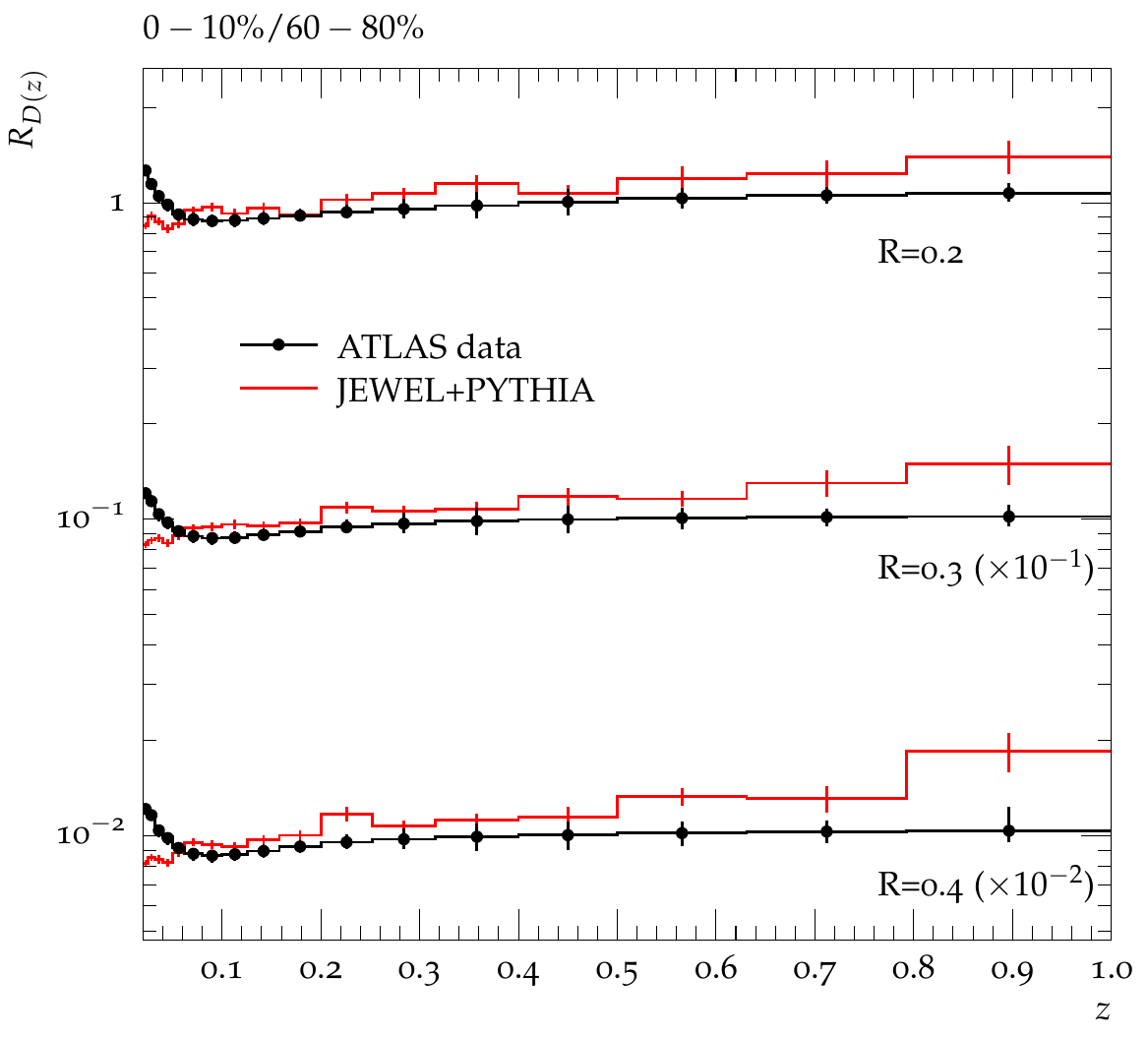}
 \includegraphics[width=0.48\textwidth]{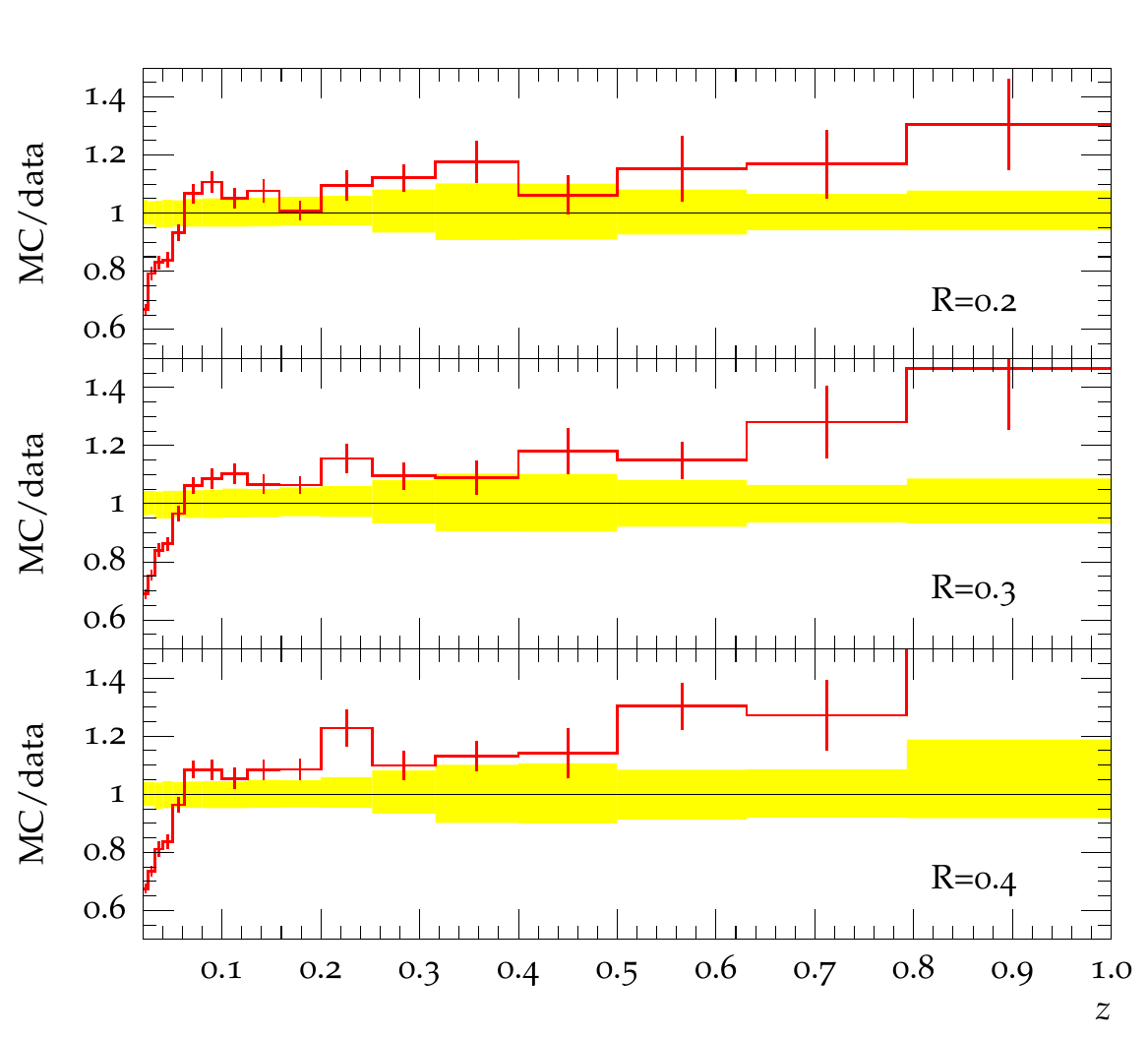}
 \caption{\textsc{Jewel+Pythia} results for the ratios of the fragmentation 
   functions $D(z)$ between central and peripheral Pb+Pb events compared to 
   \textsc{Atlas} data~\cite{ATLAS-CONF-2012-115} 
   (data points read off the plots, only maximum of statistical and 
   systematic errors shown).}
\label{Fig::RFFz}
\end{figure}

\begin{figure}[t]
 \centering
 \includegraphics[width=0.48\textwidth]{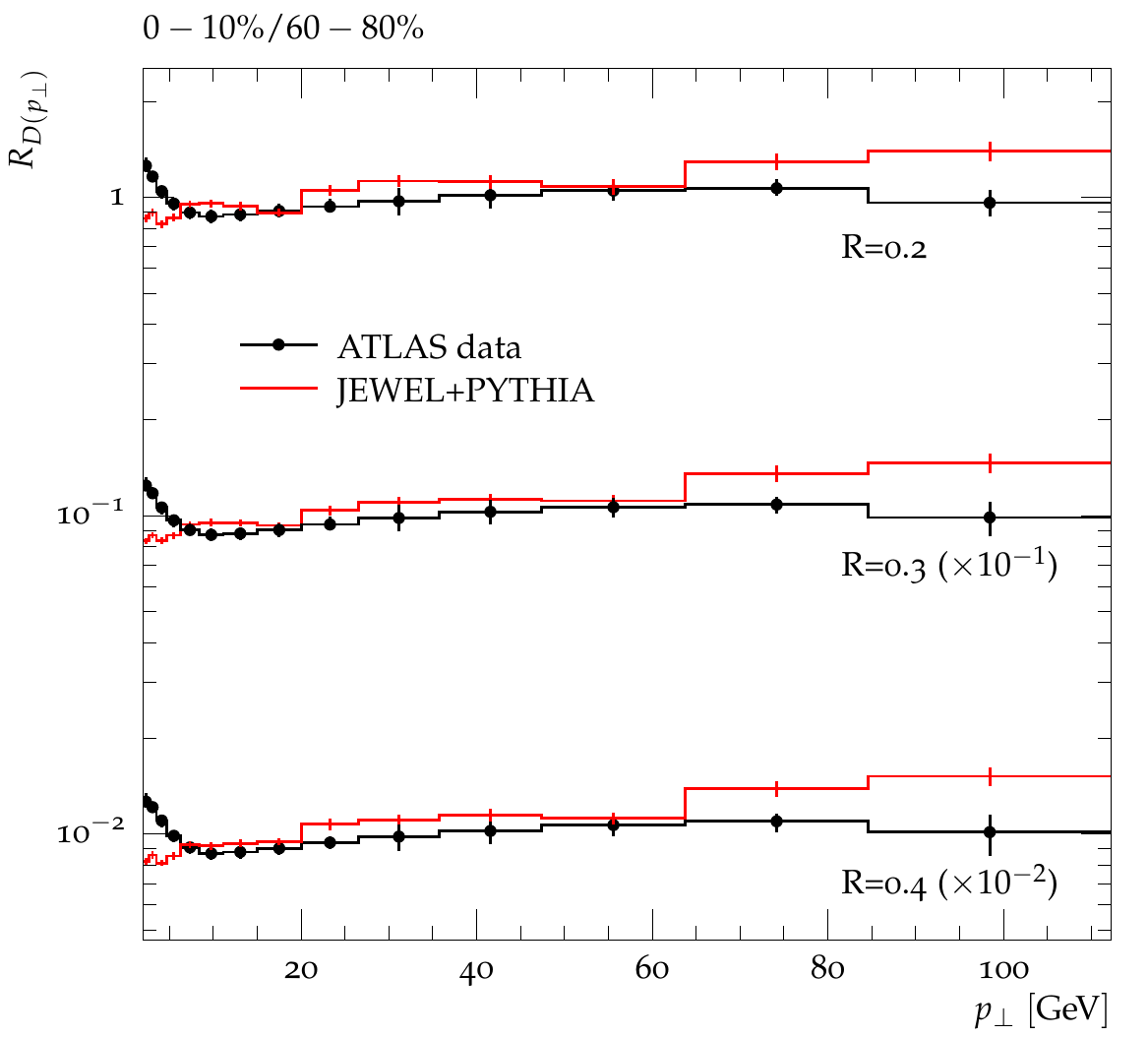}
 \includegraphics[width=0.48\textwidth]{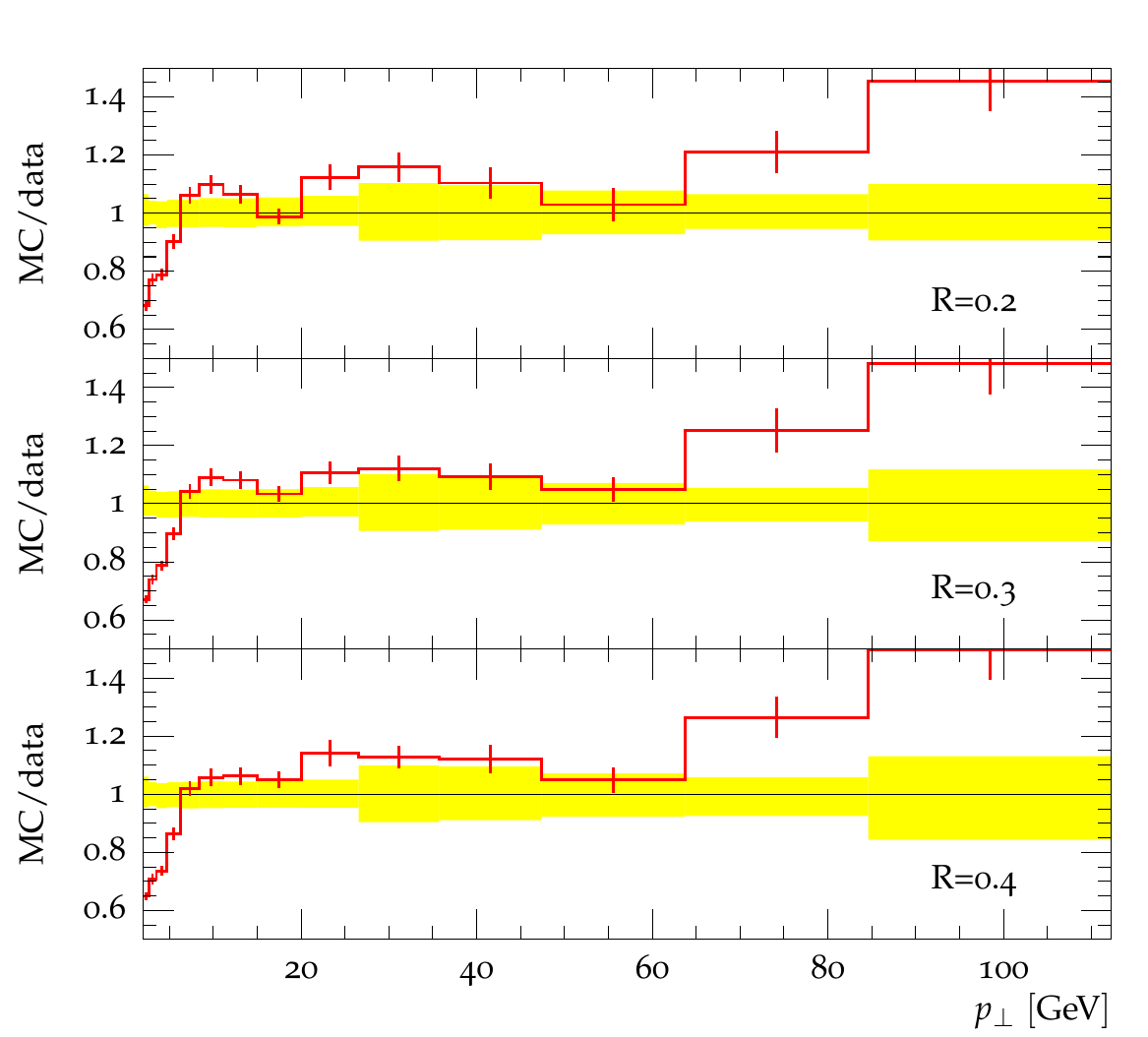}
 \caption{\textsc{Jewel+Pythia} results for the ratios of the fragmentation 
   functions $D(\pt)$ between central and peripheral Pb+Pb events compared to 
   \textsc{Atlas} data~\cite{ATLAS-CONF-2012-115} 
   (data points read off the plots, only maximum of statistical and 
   systematic errors shown).}
\label{Fig::RFFpt}
\end{figure}

The charged particle fragmentation function in jets in central Pb+Pb 
collisions as a function of the momentum fraction $z$ and relative $\pt$ in 
central and peripheral events are shown in \FigRef{Fig::FFz} and
\FigRef{Fig::FFpt}, respectively.  Except for the region of very small $z$ or 
$\pt$, which is expected to be particularly sensitive to background modelling
and subtraction, the agreement between the \textsc{Jewel+Pythia} results and 
the data is very reasonable given that for this measurement the background 
subtraction is more involved than for jets.  It remains unclear, however, 
whether the observation, that the fragmentation function in p+p at 
$\sqrt{s}=\unit[7]{TeV}$ seems to be too soft, indicates that in Pb+Pb it is 
actually too hard.  The ratio of the fragmentation functions in central and 
peripheral events shown in \FigRef{Fig::RFFz} and \FigRef{Fig::RFFpt} are 
less sensitive to such effects.  Again, the agreement between data and MC 
improves for smaller values of the jet radius.  In \textsc{Jewel+Pythia} the 
fragmentation function tends to become harder in more central events.  This 
can easily be understood since the fragmentation of the hard core remains 
unaltered by the presence of the medium while the jet energy gets degraded.

\FloatBarrier

\subsection{Uncertainties}

\begin{figure}[t]
 \centering
 \includegraphics[width=0.48\textwidth]{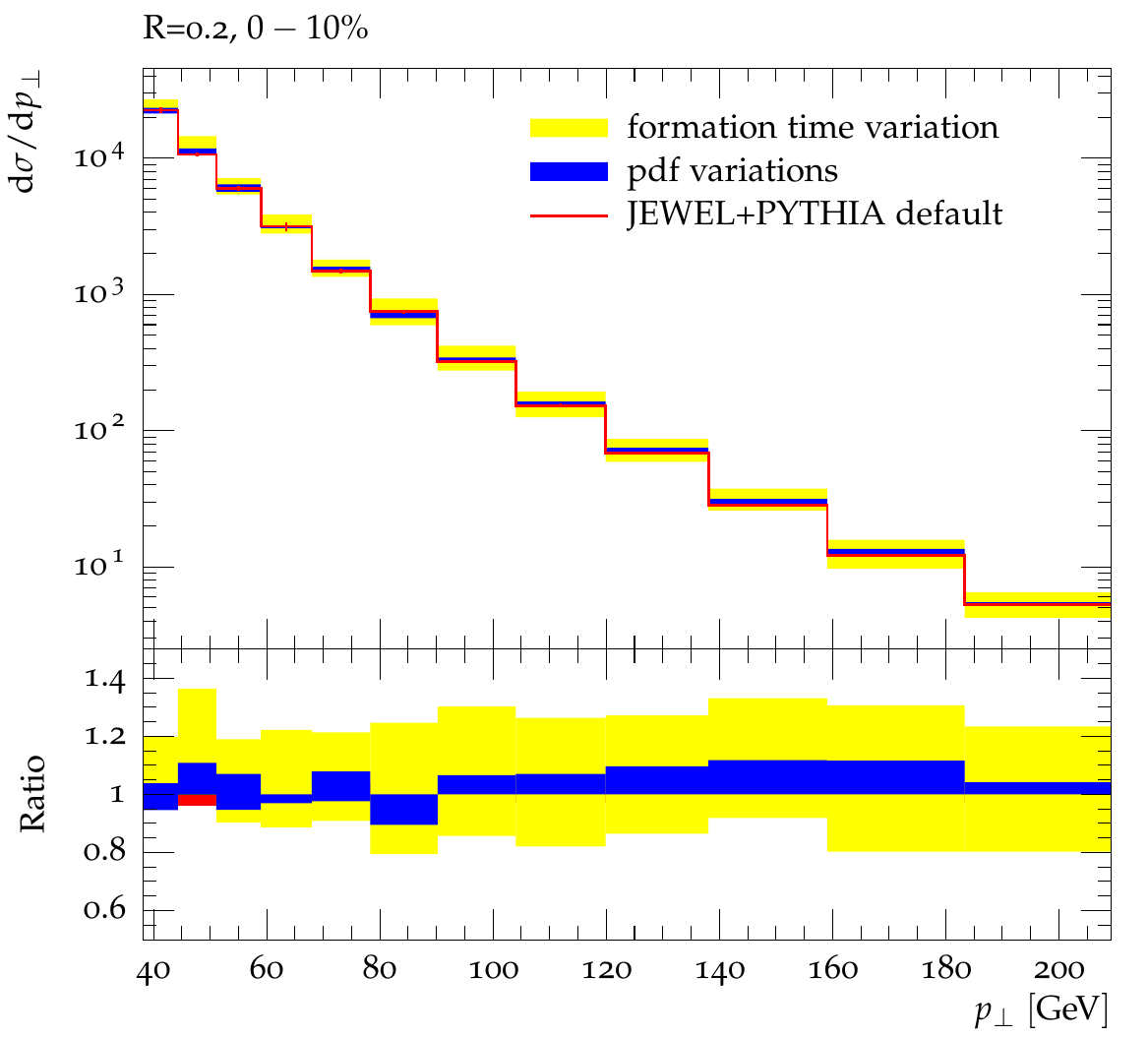}
 \includegraphics[width=0.48\textwidth]{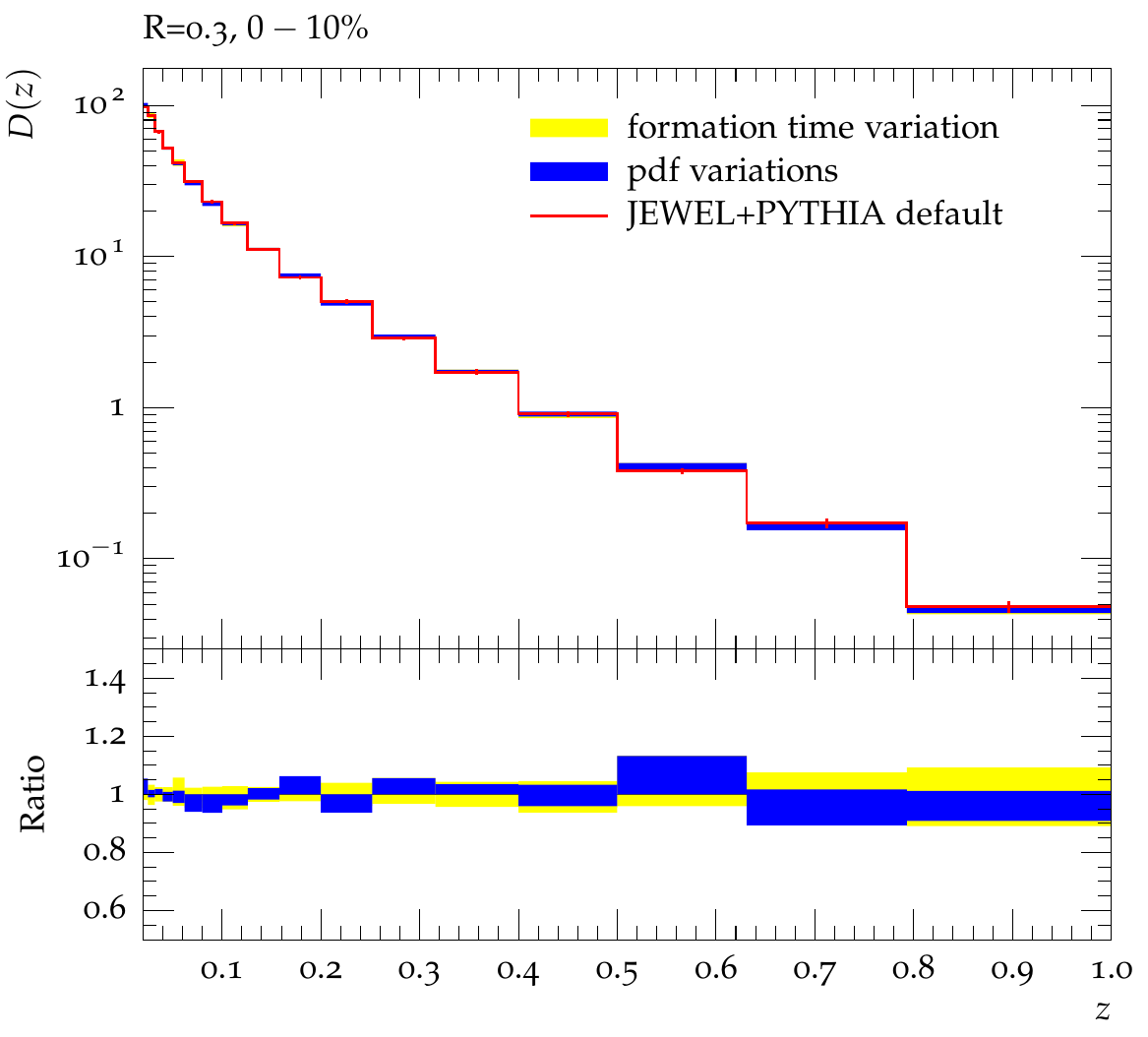}
 \caption{Uncertainties due to varying the formation time (yellow band) and 
   the pdf (blue band) on the inclusive jet spectrum (LHS) and the 
   fragmentation function (RHS) in central Pb+Pb collisions.  The statistical 
   uncertainty on the default set-up is shown as error bars in the upper 
   panel and as the red band in the ratio plots.}
\label{Fig::jetpt+FFerrors}
\end{figure}

\begin{figure}[t]
 \centering
 \includegraphics[width=0.48\textwidth]{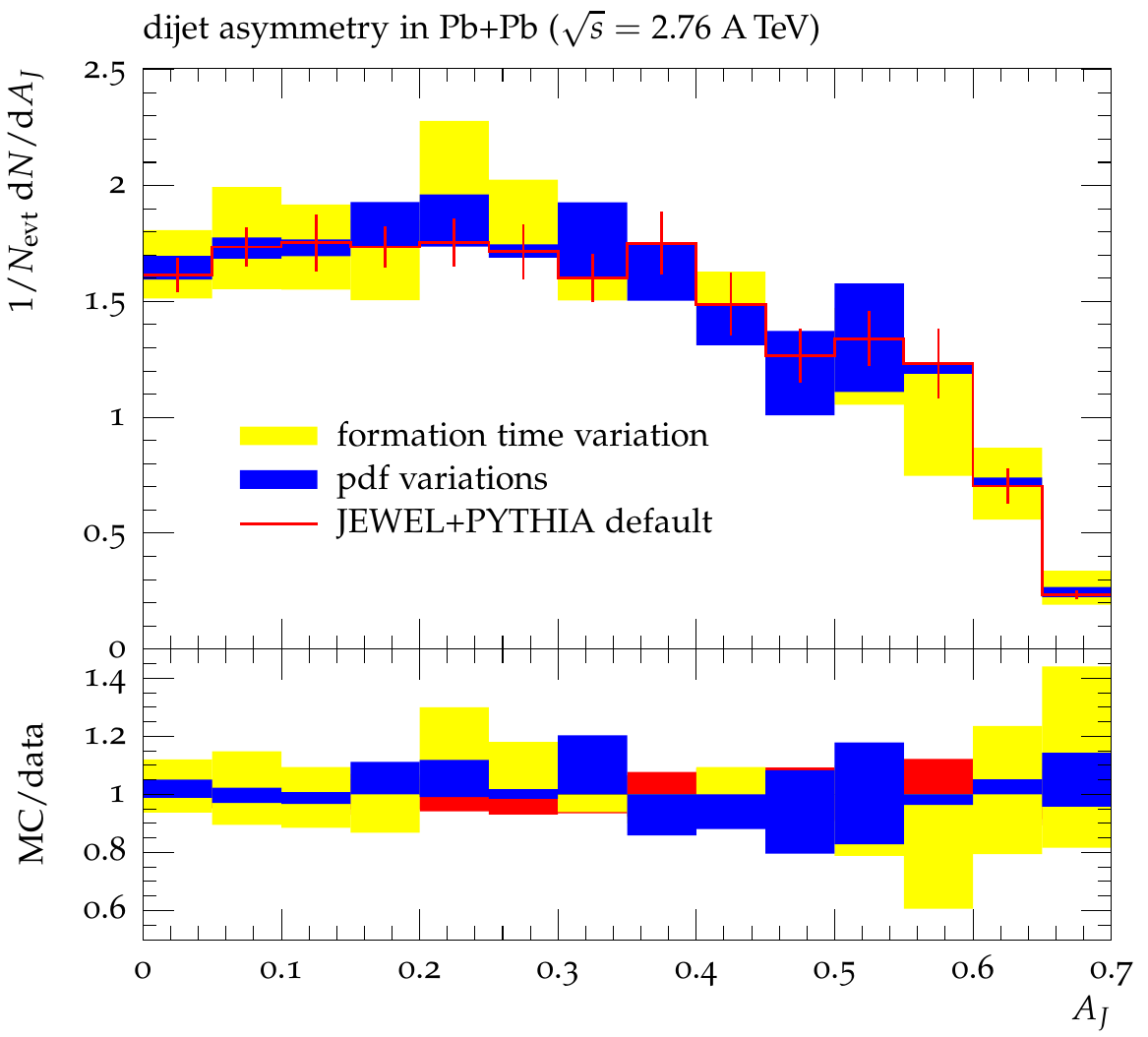}
 \includegraphics[width=0.48\textwidth]{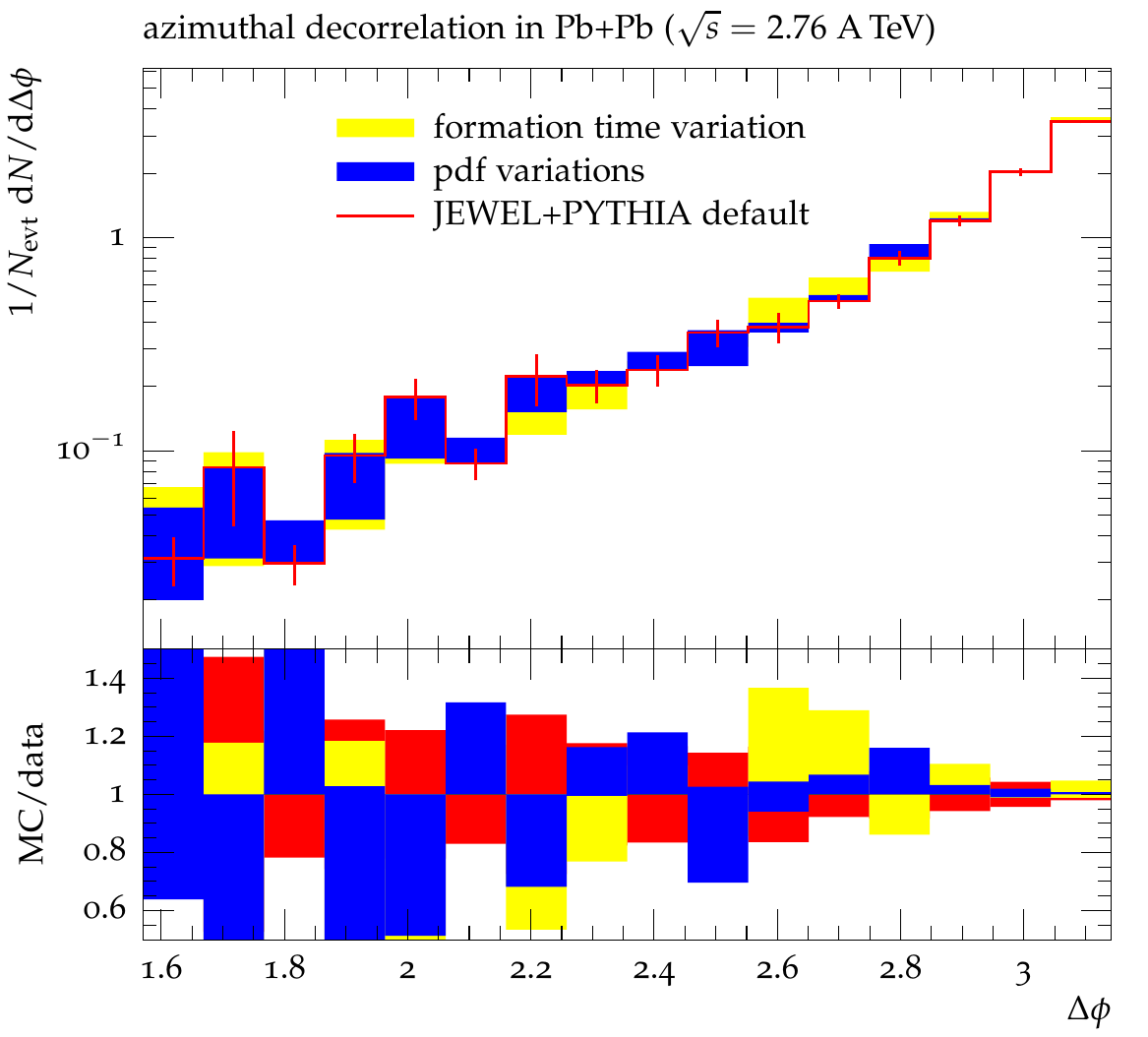}
 \caption{Uncertainties due to varying the formation time (yellow band) and 
   the pdf (blue band) on the di-jet asymmetry (LHS) and the azimuthal 
   decorrelation (RHS) in central Pb+Pb collisions. The statistical 
   uncertainty on the default set-up is shown as error bars in the upper 
   panel and as the red band in the ratio plots.}
\label{Fig::aj+decorrerrors}
\end{figure}

Uncertainties related to the basic assumptions and other aspects of the 
calculation (e.g. pdf uncertainties) are to some extent unavoidable, but can 
be quantified in our model.  While a complete assessment of all possible 
uncertainties certainly is beyond the scope of this publication, we identify 
and discuss the most important sources of uncertainties and their respective 
sizes.

Apart from the model of the medium the framework is essentially free of 
parameters. There are choices to be made concerning the infrared regulator of 
the matrix elements and the formation time, but once they are fixed they 
should not be changed. The infrared cut-off of the parton shower and the 
parameters of the hadronisation model can be tuned to LEP data and are not 
changed when going to nucleus-nucleus collisions. There is thus only very 
little room for adjustments and tuning.

The effect of varying the infrared regulator $\mu_\text{D}$, which presents 
the largest uncertainty, was already discussed in the context of the 
single--\-inclusive hadron suppression.  In this section we show the effect 
of varying the formation time up and down by a factor 2 and using different 
PDFs 
(\textsc{Cteq6l1}\cite{Pumplin:2002vw}+\textsc{Eps09lo}\cite{Eskola:2009uj}, 
which is the default, 
\textsc{Cteq6l1} without nuclear modifiactions and 
\textsc{Mstw08Lo}\cite{Martin:2009iq}+\textsc{Eps09lo}). 
Figure~\ref{Fig::jetpt+FFerrors} shows the effect on the inclusive jet spectrum 
and the fragmentation function and \FigRef{Fig::aj+decorrerrors} the di-jet 
asymmetry and azimuthal decorrelation. Varying the formation times leads to a 
$\sim \unit[20]{\%}$ variation of the jet rate in central Pb+Pb collisions 
and a smaller but visible change in the di-jet asymmetry and azimuthal decorrelation while the fragmentation function is 
fairly insensitive.  The uncertainty related to the choice of PDF is smaller 
and of the same size as the current statistical uncertainty on the MC results.

In a similar way practically all aspects of the simulation can be varied and 
the uncertainties estimated.  We find that the largest sources of 
uncertainties are the choice of the infrared regulator and to a lesser degree 
the formation time.

\section{Conclusions}
\label{Sec::Conc}

In this paper we report on the development of a novel description of jet 
quenching and its implementation into the Monte Carlo generator \textsc{Jewel}. 
The paradigm underlying the construction of our model is the language given 
by perturbative QCD, with the aim to enable a quantitative discussion of 
effects beyond it.  In this context, hard interactions of a jet with the 
medium resolve its constituents as quasi-free partons and should be described 
in perturbation theory\footnote{In general, the scale above which the
components of a final state parton shower can resolve partonic
constituents of the medium as quasi-free partons depends on properties of
the medium~\cite{D'Eramo:2012jh}.
Comparing a perturbative formulation of jet quenching to data
can thus give access to the nature of medium constituents.}.  
In reinterpreting induced radiation as regular, 
although infra-red-regulated, $2\to 2$ parton scatterings, supplemented with 
a parton shower we arrive at a framework whose dynamics is entirely based on 
standard perturbative technology also used in the simulation of $p+p$ 
collisions and which is minimal in its assumptions.  It is formulated entirely 
in general kinematics\footnote{Although the present code goes far beyond 
  the eikonal kinematics, it is clear how to recover this limit of the full 
  emission pattern encoded here.} 
and thus overcomes limitations of previous approaches which were entirely 
based on results obtained in eikonal or close-to-eikonal kinematics~\cite{Armesto:2011ht}. 
Consequently, there is no distinction between elastic and inelastic scattering.
Rather, both are possible outcomes of the same process and the parton shower 
cut-off defines the separation between the two.  Also the distinction between 
vacuum and medium--\-induced radiation has become obsolete, as they are 
fundamentally the same.  The interplay between all sources of radiation is 
governed by the formation times and generally a certain emission cannot be 
classified as vacuum-like or medium-induced in a meaningful way. 

The emerging framework is well constrained by data from elementary reactions 
and does not leave much room for tuning.  Apart from the model of the medium 
there is only a single quantity with a significant uncertainty that should 
perhaps be regarded as a parameter, namely the infrared regulator of the 
matrix elements.  All uncertainties related to assumptions or other aspects 
of the calculations can be quantified, examples for the numerically most 
important uncertainties have been presented here, but a complete 
characterisation of all possible uncertainties is beyond the scope of this 
publication. 

Our new framework does not make any assumptions about the medium other than 
that is partonic.  The phase space distribution does not matter and in 
principle any distribution can be interfaced.  The results presented here 
were obtained with a simple Bjorken type model.  Nevertheless, 
after fixing the parameters with \textsc{RHIC} data, the agreement with very 
different jet and leading hadron measurements at the \textsc{LHC} is very 
reassuring.  This leads us to conclude that our calculation, although certain 
aspects are simplified, captures the relevant physics underlying jet quenching.
At the same time the overall good description of data lets us speculate in how
far more complicated medium models may obfuscate simple physical realities.
This clearly is a question that could only be answered seriously by interfacing
more medium models and cataloguing the results obtained with them.

However, we would like to stress that for some observables there appear to
be unresolved and potentially unresolvalve issues with the subtraction of
backgrounds, where the procedure in the analysis of MC events differs from 
the experimental one and may lead to discrepancies.  The observation that 
generally the agreement between MC and data gets worse for larger jet radii 
points in this direction. Also, it should be noted that there is a class of 
measurements that are sensitive to the back-reaction of the jet on the medium 
and that therefore our calculation cannot be expected to describe.  This
problem could only be overcome by trying to extend our model to the fully
inclusive treatment of the full collision -- something clearly beyond the
scope of this paper.

Apart from the obvious option of testing different models for the medium we 
have hinted at various other ways of how our calculation can systematically be 
improved using a well--\-understood perturbative language.

\bibliographystyle{JHEP}
\bibliography{jetquenching}

\end{document}